# Review of Newman's Analytical Series on Disk Electrodes

Boshuo Wang [1, *] and James D. Weiland [2, 3]

[1] Department of Psychiatry and Behavioral Sciences, School of Medicine, Duke University, Durham, NC 27710, USA

[2] Department of Biomedical Engineering, College of Engineering and Medical School, and [3] Department of Ophthalmology and Visual Sciences, Medical School, University of Michigan, Ann Arbor, MI 48109, USA

[*] Author to whom correspondence should be sent to: boshuo.wang@duke.edu.

**Keywords**—disk electrode, electrode–electrolyte interface, double layer capacitance, charge transfer resistance, ohmic resistance; frequency dispersion, current step response, voltage step response, steady state solution, transient solution; analytical solution, spectral analysis, eigenvalues, rotational elliptic coordinates, Legendre function; numeric solution, numeric accuracy.

**Abstract**

Dr. John S. Newman, an expert and pioneer in electrochemical engineering, studied the electrical characteristics of disk electrodes extensively since the 1960s. Newman and his colleagues published the results in a series of articles in the *Journal of the Electrochemical Society*. This seminal series is consistent and well-written and has been cited by many in electrochemistry and closely related fields. However, the articles, especially the later ones in the series, enjoined less familiarity in other fields, including biomedical engineering in which electrodes became widely used in neural stimulation. The purpose of this review is therefore to summarize Newman's work on disk electrodes together and provide a comprehensive understanding of the original articles. The review mainly focuses on the behaviors of interest to neural stimulation, namely the primary distribution, frequency dispersion, and the current step and voltage step responses. More mathematical details are supplemented to the original calculation to help the readers follow the derivation more easily. Several adjustments are made to Newman's original analyses. First, the equation sets are summarized into matrix form, which demonstrates the underlying structure of the electrode-electrolyte system. This formulation is helpful in showing the similarity and differences between the different responses. Also, the normalization factors to give dimensionless variables have been slightly scaled by $\pi/4$ compared to the original articles, which endows them the representation of physical quantities. A consistent symbol naming system is used to refer to the results across different articles. Finally, some preliminary analyses are presented on the numeric accuracy of the solutions. The review will provide a comprehensive understanding of the original articles, especially in the context of neuroengineering applications.

**Acknowledgments**

The first version of this review was completed in early 2012 at the University of Southern California (USC) as a study report. In 2013, a shortened version only discussing the primary distribution (Newman, 1966a) and current step response (Nisancioğlu and Newman 1973a) was planned as supplementary materials for the article Wang et al., (2014); it was ultimately not submitted due to the revision of the manuscript shortening the length and reducing significant amount of discussion on the analytical framework by Nisancioğlu and Newman (1973a), and also due to the journal's policy of not allowing online supplementary material. The review was further revised in 2015 and subsequently included as Chapter 6—Supplementary Materials in B. Wang's dissertation, Investigation of the Electrode-Tissue Interface of Retinal Prostheses, Department of Biomedical Engineering, USC, Los Angeles, CA, USA, May 2016, ProQuest Dissertation No. 10124439. The authors acknowledge support from USC under a Viterbi Fellowship, the Biomimetic MicroElectronic Systems Engineering Research Center (BMES ERC) of the National Science Foundation (NSF) under Grant EEC-0310723, and the National Institutes of Health (NIH) under Grant U01 GM104604.

This current version includes an expanded introduction, reorganization of the contents, correction of mathematical terminology, improved solution methods for the eigenvalue problems, additional analysis and details, and editing of the text and remaking of figures. The MATLAB code and data related to this review have been made available online at GitHub: 10.5281/zenodo.4291332

Critical comments on the contents and suggestions for improvement are most welcome!

# Introduction

Dr. John S. Newman, an expert and pioneer in electrochemical engineering, studied the electrical characteristics of disk electrodes extensively since the 1960s. By applying his ability to reduce complex problems to their essential core elements and mastery of mathematical analysis (Newman and Battaglia, 2018), the system of equations describing the disk electrodes was solved analytically with ease and elegance. The results were published by Newman and his colleagues in the *Journal of the Electrochemical Society* and the articles formed a seminal series on this topic, which was well-written, consistent, and cited by many in electrochemistry, electrochemical engineering, and closely related fields. In the field of biomedical engineering, electrodes have become widely used in neural engineering applications. While the first article in Newman's series on the electrolyte's access resistance of the disk electrode [1] is well known, the series as a whole, and especially the later articles, enjoined less familiarity in the biomedical fields. Not only were many recent studies not placed into the context of these later publications by Newman et al., the wheels were sometimes completely reinvented with modern numeric and analytical methods.

The purpose of this review is therefore to summarize Newman's work on disk electrodes together and provide a comprehensive understanding of the original articles, which are listed and introduced in the next section. The review mainly focuses on the behaviors of interest to neural stimulation, namely the Primary Distribution [1], Frequency Dispersion [3] and the Current Step [4] and Voltage Step [5] responses. The primary distribution is determined by the ohmic resistance of the electrolyte; it occurs at the very beginning of a pulse when a voltage or current input is applied to the electrode and serves as the basis for other types of responses. The frequency dispersion is relevant for alternating current (AC) stimulation, such as transcranial AC stimulation (tACS) and kHz stimulation for nerve block. The step responses are relevant for transcranial direct current stimulation (tDCS) and the most commonly used form of stimulation using current- or voltage-controlled rectangular pulses, which are superposition of step inputs with various delays and amplitudes; responses to arbitrarily-shaped pulse waveforms can also be obtained from the step response by convolution (Wang et al., 2014).

Disk electrodes are used differently in electrochemistry and neuroengineering applications, such as having specific geometric configurations, being fabricated with distinct materials, and interfacing with electrolyte of different properties. As the titles of many articles in the series state, disk electrodes are often rotated in electrochemical engineering studies, which establishes a steady field of convection in the electrolyte to support relatively fast electrochemical reactions that would otherwise be limited by diffusion alone. Electrodes in neuroengineering, however, emphasize foremost safety and biocompatibility, and thus avoid movements and electrochemical reactions as much as possible. The inert materials, small amplitudes of electrode polarization, and short pulse widths or low pulse rates of many stimulation paradigms, however, prove ideal for placing neuroengineering electrodes into Newman's framework in which the diffusion layer is neglected and the interface is linearized for deriving analytical solutions, as is further discussed below in the comments on [2] and when the electrode-electrolyte model is introduced and solved. The implanted microelectrodes also interface with limited electrolyte space and thus the environment is quite different compared to electrochemical cells. However, the behavior of the interface is dominated by the electrolyte in the vicinity of the electrode surface, and due the contrast between the less conductive surrounding tissue and the conductive fluids with which the electrodes typically interfaces directly, such as the perilymph, vitreous humor, and cerebral spinal fluid, many of the ideal geometric assumptions in the analytical framework can be translated to realistic situations with fairly good accuracy.

Several adjustments and additions are made to Newman's original analysis. First, the equation sets are summarized into matrix form, which demonstrates the underlying structure of the electrode-electrolyte system, placing the problem in the context of spectral analysis. This formulation is also helpful in showing the similarity and differences between the three different inputs analyzed in this review, i.e., sinusoidal voltage input, current step input, and voltage step input, and also provides easier implementation with the help of nowadays computer programs, e.g. MATLAB. Further, the Normalization Factors to give dimensionless variables are defined based on quantities of the primary distribution and therefore have been slightly scaled by $\pi/4 \approx 0.785$ compared to the original articles. This scaling endows the factors the representation of physical quantities, instead of mere normalization purpose. Typical values of the electrode parameters and the normalization factors are given as well, which are presented and discussed following the results in the Primary Distribution section where they are first introduced. Further, a consistent variable and symbol naming system is used to refer to the results across different articles, as given in the Symbol Naming section. Finally, some preliminary analyses are presented on the Numeric Accuracy of the solutions in the broader context of spectral analysis, which include the eigenvalue problem and the spatial distributions of the current density.

This review is suitable for anyone interested in the electrode–electrolyte interface, especially in the context of neuroengineering applications. Dr. Newman's original derivation was quite concise and omitted many details. We supplemented a significant amount of mathematical calculation to help the readers follow the step-by-step derivation more easily and a mathematical appendix on Legendre functions is provided, thus making this review accessible even to undergraduate students with only intermediate level knowledge in partial differential equations and the relevant physics and chemistry. A numeric appendix containing the solutions to the system of equations is provided. Whereas the presented solutions are specific to the disk electrode in an ideal situation, the principles, such as the decomposition of the spatial components of the solution using orthonormal basis functions (Chen et al., 2020) and separation of the temporal component into steady state response and transient response, are generalizable to many situations, such as electrodes of other geometry, with protrusion from or recession into the substrate, limited and irregular electrolyte space, and/or nonlinear interface parameters. These principles can be applied to or utilized in solving and optimizing electrode interfaces, before falling back on methods to perform brute force or "intelligent" searches to obtain numeric solutions. After all, "computation is temptation that should be resisted as long as possible" (Boyd, 2001).

**List of Articles in the Series**

Whereas other articles on disk electrodes by Newman and colleagues exists, only those published in the *Journal of the Electrochemical Society* are included here. This is not meant to be an exhaustive list, as some articles on the convective flow or other topics not so closely related to this review are not discussed. The articles directly covered by this review are numbered, whereas other articles are listed by bullet points.

[1] *Resistance for Flow of Current to a Disk*, vol. 113, no. 5, pp. 501–502, 1966a. This is the 1st article on this topic and gives the steady state solution to an ideal disk electrode without considering overpotentials on the electrode surface related to the double layer capacitance, Faradaic reactions, or diffusion. The solution is the primary current density distribution of the current/voltage step input. The voltage-current relationship gives the resistance of the electrolyte that can be obtained experimentally using the interrupter technique and provides the normalization factors for the following problems. The rotational elliptic coordinates are introduced to solve Laplace's equation, however due to the simplicity of the primary distribution, few details are given on how to solve the partial differential equations. This article's steady rate of citation and high citation volume (832 as of November 2020) was discussed by Newman's colleagues Orazem and Tribollet (2009).

[2] *Current Distribution on a Rotating Disk below the Limiting Current*, vol. 113, no. 12, pp. 1235–1241, 1966b. This article addresses how the diffusion layer may shape the current density profile, when the charge transfer reaction but not the double layer charging is considered. Two limiting situations are discussed, namely the current density on the disk when the potential just outside the diffusion layer of the disk surface is uniform, i.e., the primary current distribution determined by ohmic resistance only, and, vice versa, the potential distribution when the current density is uniform. If the current density on the electrode is small compared to the limiting current density determined by mass transport, the concentration overpotential can be neglected and both the ohmic and kinetic resistances determine the secondary current distribution. Further including the mass transport resistance results in the tertiary distribution. Assuming the current density does not become limited by the diffusion layer, its effect is not studied in many later articles in the series, especially those of interest for this review. This article gives more detail on how to solve Laplace's equation in rotational elliptic coordinates, including crucial steps for understanding the calculations in later articles.

- *The Diffusion Layer on a Rotating Disk Electrode*, vol. 114, no. 3, p. 239, 1967. A brief follow-up discussion on numeric methods for solutions of the diffusion layer in [2].

- *Current Distribution on a Rotating Disk*, with V. Marathe, vol. 116, no. 12, pp. 1704–1707, 1969. Summary and experimental verification of [2].

[3] *Frequency Dispersion in Capacity Measurements at a Disk Electrode*, vol. 117, no. 2, pp. 198–203, 1970a. This articles studies sinusoidal voltage input to the disk electrode, and it includes the double

layer capacitance and a linearized Faradaic reaction. The effective series resistance and electrode capacitance are calculated as a function of frequency, and include the influence of the charge transfer resistance, which is not modeled explicitly. More details on numeric solution of the Laplace equation are given, which, compared with the following two situations, will show consistency in the solutions when presented in a matrix format.

- *Ohmic Potential Measured by Interrupter Techniques*, vol. 117, no. 4, pp. 507–508, 1970b. The ohmic potential measured by current interruption, i.e., current steps, correspond to the primary current distribution. A discussion on the time constants involved in the discharging of the double layer capacitance is provided.

- *Limiting Current on a Rotating Disk with Radial Diffusion*, with W. H. Smyrl, vol. 118, no. 7, pp. 1079–1081, 1971. Discussion on the effect when diffusion in the radial direction is considered.

- *Detection of Nonuniform Current Distribution on a Disk Electrode*, with W. H. Smyrl, vol. 119, no. 2, pp. 208–212, 1972. Discussion on the implication of the nonuniform current distribution and its detection and measurement.

- *The Error in Measurements of Electrode Kinetics Caused by Nonuniform Ohmic-Potential Drop to a Disk Electrode*, with W. H. Tiedemann and D. N. Bennion, vol. 120, no. 2, pp. 256–258, 1973. Discussion on the consequence of the nonuniform potential distribution in the electrolyte above the electrode surface on the measurements of electrode parameters.

[4] *The Transient Response of a Disk Electrode*, with K. Nisancioğlu, vol. 120, no. 10, pp. 1339–1346, 1973a. This article describes the response to a current step input and includes the double layer capacitance and a linearized Faradaic reaction. The method decomposes the response into a steady state response and a transient response, the latter being an eigenvalue problem consisting of eigensolutions associated with the different time constants.

[5] *The Transient Response of a Disk Electrode with Controlled Potential*, with K. Nisancioğlu, vol. 120, no. 10, pp. 1356–1358, 1973b. This article describes the response to a voltage step input, using similar technique as described in [4]. The study discusses the different time constant associated with the transient response, especially the zeroth order time constant that is unique to the voltage step input.

- *Corrosion of an Iron Rotating Disk*, with N. Vahdat, vol. 120, no. 12, pp. 1682–1686, 1973. Application of the disk electrode model to calculate corrosion rate on iron disk electrodes.

- *An Asymptotic Solution for the Warburg Impedance of a Rotating Disk Electrode*, with R. V. Homsy, vol. 121, no. 4, pp. 521–523, 1974a. Analytical approximation of the Warburg impedance that models the diffusion at the interfaces at high frequencies. Related to [2].

- *The Short-Time Response of a Disk Electrode*, with K. Nisancioğlu, vol. 121, no. 14, pp. 523–527, 1974. Since the transient responses in [4] and [5] involve an infinite set of eigenfunctions, each being

- a combination of the infinite set of basis functions, the numeric solution can be very laborious to solve. This article provides an efficient asymptotic solution for the short-time transients.

- *Current Distribution on a Plane below a Rotating Disk*, with R. V. Homsy, vol. 121, no. 11, pp. 1448–1451, 1974b. A stationary disk electrode is placed below the rotating disk. The solution of Laplace's equation in rotational elliptic coordinates appears in the electrolyte layer immediately above the electrode.

- *Current Distribution on a Disk Electrode for Redox Reactions*, with P. Pierini and P. Appel, vol. 123, no. 3, pp. 366–369, 1976. Development of the models for the overpotential and diffusion layer. The solution of Laplace's equation in rotational elliptic coordinates appears in the electrolyte layer immediately above the electrode.

- *Potential Distribution for Disk Electrodes in Axisymmetric Cylindrical Cells*, with P. Pierini, vol. 126, no. 8, pp. 1348–1352, 1979. Solution for disk electrode with limited electrolyte space.

- *Analytic Expression of the Warburg Impedance for a Rotating Disk Electrode*, with B. Tribollet, vol. 130, no. 4, pp. 822–824, 1983. Analytical expression of the Warburg impedance. Related to previous work by Homsy and Newman (1974a).

- *Corrosion of a Rotating Iron Disk in Laminar, Transition, and Fully Developed Turbulent Flow*, with C. G. Law, vol. 133, no. 1, pp. 37–42, 1986. The calculation of the potential distribution utilizes the techniques and results of earlier work in the series.

- *The Kramers-Kronig Relations and Evaluation of Impedance for a Disk Electrode*, with M. M. Jakšić, vol. 133, no. 6, pp. 1097–1101, 1986. The impedance spectroscopy of electrodes obeys the Kramers-Kronig relation. Based on this relation, the capacitance as a function of frequency [3] can be calculated from the effective electrode resistance, and vice versa.

- *Current Distribution at Electrode Edges at High Current Densities*, with W. H. Smyrl, vol. 136, no. 1, pp. 132–139, 1989. Calculation of the large current density at the electrode's edge for Tafel kinetics. Related to previous work by Nisancioğlu and Newman (1974).

- *Corrections to Kinetic Measurements Taken on a Disk Electrode*, with A. C. West, vol. 136, no. 1, pp. 139–143, 1989. Correcting the errors due to the nonuniform current distribution of the electrode and showing the condition, under which the errors can be neglected. Related to previous work by Tiedemann, Newman, and Bennion (1973).

- *Current Distribution near an Electrode Edge as a Primary Distribution Is Approached*, with A. C. West, vol. 136, no. 10, pp. 2935–2939, 1989. Current density at the electrode's edge when the insulator and electrode have arbitrary angle of intersection, of which the disk electrode is a special case. Related to previous work by Nisancioğlu and Newman (1974) and Smyrl and Newman (1989).

- *Cathodic Protection for Disks of Various Diameters*, with S. X.-Z. Li, vol. 148, no. 4, pp. B157–B162, 2001. Applying the disk electrode model to cathodic protection.

# Model of the Disk Electrode

## System of Equations

A disk electrode of radius $r_0$ is embedded in an infinite large insulating substrate interfacing with a semi-infinite large space of electrolyte of isotropic conductivity $\kappa$. Voltage or current is applied to the metal part of the electrode, which is equipotential, and the ground is located at infinity. A cylindrical coordinate system $(r, \phi, z)$ is established with the origin at the disk's center and the $z$ axis pointing perpendicular into the electrolyte space (Figure 1).

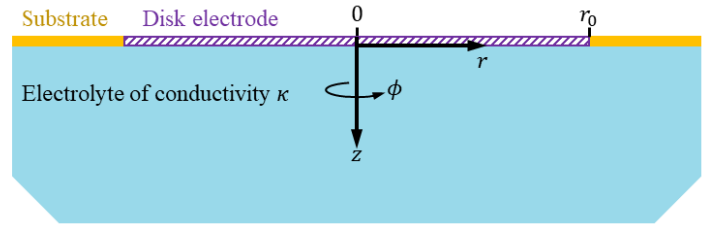

Figure 1. Illustration of disk electrode and the cylindrical coordinate system. Adapted from Wiley and Webster (1982).

As the bulk electrolyte space contains no sources or sinks and no concentration gradients of the solvents, current continuity yields

$$\nabla \cdot \boldsymbol{J} = 0 \quad , \quad r, z \geq 0 \quad , \tag{1}$$

for the current density $\boldsymbol{J}$. Together with Ohm's law

$$\boldsymbol{J} = -\kappa \nabla \varphi = -\kappa \left( \frac{\partial \varphi}{\partial r} \hat{\boldsymbol{r}} + \frac{1}{r} \frac{\partial \varphi}{\partial \phi} \widehat{\boldsymbol{\phi}} + \frac{\partial \varphi}{\partial z} \hat{\boldsymbol{z}} \right) \quad , \tag{2}$$

where $\hat{\boldsymbol{r}}$, $\widehat{\boldsymbol{\phi}}$, and $\hat{\boldsymbol{z}}$ are unit vectors of the coordinate system, Laplace's equation holds for the electric potential $\varphi(r, \phi, z, t)$

$$\nabla^2 \varphi = \frac{1}{r} \frac{\partial}{\partial r} \left( r \frac{\partial \varphi}{\partial r} \right) + \frac{1}{r^2} \frac{\partial^2 \varphi}{\partial \phi^2} + \frac{\partial^2 \varphi}{\partial z^2} = 0 \quad , \quad r, z \geq 0 \quad . \tag{3}$$

The second term is always zero as the system is axisymmetric, and the azimuth $\phi$ is omitted in all further analysis. The time variable $t$ is only specifically shown when its inclusion is necessary for disambiguation.

General boundary conditions for the voltage and current that apply to all the situations studied include

$$\begin{cases} |\varphi(r,z)| < +\infty \quad , \quad r \geq 0, z \geq 0 \\ \varphi(r,z) = 0 \quad , \quad r, z \to +\infty \end{cases} \tag{4}$$

$$\begin{cases} J_z(r,z) = J_0(r) \quad , \quad z = 0^+, r \leq r_0 \\ J_z(r,z) = 0 \quad , \quad z = 0, r > r_0 \end{cases} \tag{5}$$

where $J_0(r)$ and $J_z(r, 0^+)$ denote the normal current density trough the electrode surface and axial current density immediately outside the diffusion layer in the bulk electrolyte (Figure 2). These conditions state that, the potential within the electrolyte is finite, the ground is at infinity, the current density on the electrode surface is continuous with that in the electrolyte, and the substrate is insulating, respectively.

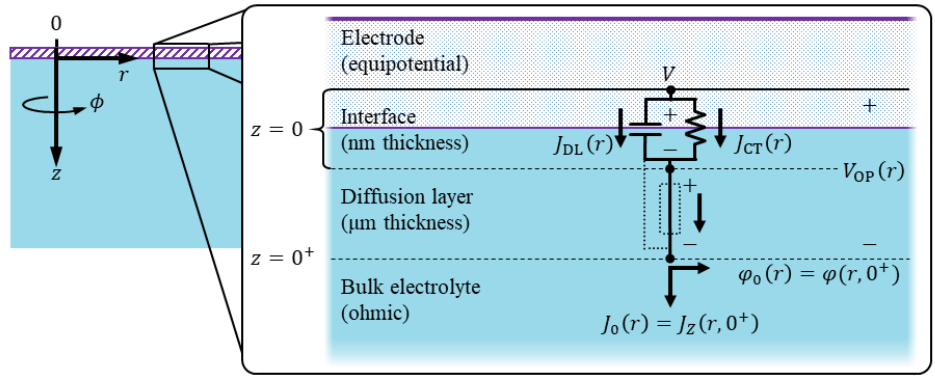

Figure 2. The current density components within the microscopic interface of the electrode consists of capacitive and Faradaic components. The diffusion layer has negligible thickness compared to the dimension of the bulk electrolyte and the current density within this layer is typically considered to be normal to the electrode surface and contains no tangential components. The concentration gradients within the diffusion layer can be represented either as a concentration overpotential or as an impedance (dashed outline). In the Randles circuit, the Warburg element associated with the concentration polarization is typically connected in series with the charge transfer resistance, and the interface's double layer is connected in parallel with these two elements (dashed line) and often represented as a constant phase element. The current density is continuous through the interface and diffusion layer and with the normal current density of the bulk electrolyte at the outer border of the diffusion layer.

The above boundary conditions do not directly connect the potential $\varphi(r,z)$ and current $\mathbf{J}(r,z)$ in the bulk electrolyte with the voltage $V$ or current $I$ input applied to the electrode. To complete the description of the disk electrode, the electrode voltage $V$ is related to the bulk electrolyte potential right outside the diffusion layer $\varphi_0(r) = \varphi(r, 0^+)$. The difference between the two is defined as the overpotential

$$V_{\text{OP}}(r) = V - \varphi_0(r) \quad , \quad r \leq r_0 \quad . \tag{6}$$

The overpotential can be considered to consist of two components (Figure 2). A surface/activation overpotential spans across the microscopic interface, within which the double layer charging and Faradaic charge transfer occur. Between the outer boundary of the interface and the inner boundary of the bulk electrolyte, concentration gradients may exist in a thin diffusion layer due to mass transport not being able to replenish or diffuse reactants fast enough to and away from the electrode surface. Due to the thinness of this layer, diffusion and the associated current density is assumed to be only in the normal direction of the electrode surface. Depending on the formalism of the interface model, the influence of the concentration on the potentials can be either described as a concentration overpotential or an equivalent circuit element, such as the Warburg element associated with charge transfer or as a constant phase element to combine its effect with the double layer (Newman, 2004; Orazem and Tribollet, 2008). Assuming that the current densities across the electrode are small and not limited by mass transport, the concentration gradients are ignored and so is the associated overpotential in the diffusion layer.

The current density due to the charging or discharging of the double layer is described by a differential capacitance

$$J_{\text{DL}}(r) = \gamma \frac{\partial(V_{\text{OP}}(r))}{\partial t} \quad , \quad r \leq r_0 \quad , \tag{7}$$

with $\gamma$ being the double layer capacitance per unit area. The double layer capacitance is voltage dependent and non-linear, deserving a dedicated book chapter to fully discuss its behavior (Newman, 2004), but for simplicity is considered constant in the subsequent analysis.

Under small surface overpotential $V_{\text{OP}}$ (defined relative to an equilibrium potential), the current density of the charge transfer reaction can be given by a linear approximation

$$J_{\text{CT}}(r) = g_{\text{CT}} \cdot V_{\text{OP}}(r) \quad , \quad r \leq r_0 \quad , \tag{8}$$

where the charge transfer conductance

$$g_{\text{CT}} = (\alpha_a + \alpha_c)\frac{j_0 ZF}{RT} \tag{9}$$

relates to the kinetics of a single Faradaic reaction described by the Butler–Volmer equation. Here, the parameters $\alpha_a$ and $\alpha_c$ are the anodic and cathodic charge transfer coefficients, respectively, $j_0$ is the exchange current density, $Z$ is the number of electrons involved in the reaction, $T$ is the absolute temperature, $F$ is the Faraday constant, and $R$ is the universal gas constant. Again, this approximation is a significant oversimplification of the complex behavior of the charge transfer process (Newman, 2004), however, is often sufficient for the chemically-inert electrodes in biomedical applications.

Together, the current densities of the double layer and charge transfer components are part of the boundary condition of current continuity

$$J_0(r) = J_{\text{DL}}(r) + J_{\text{CT}}(r) = J_z(r,z) \quad , \quad z = 0^+, r \leq r_0 \quad , \tag{10}$$

Combining (6)–(8) into Eq. (10), the electrode voltage $V$ and the electrolyte potential $\varphi_0(r)$ are therefore related by

$$\gamma \frac{\partial(V - \varphi_0(r))}{\partial t} + g_{\text{CT}}(V - \varphi_0(r)) = -\kappa \frac{\partial \varphi(r,z)}{\partial z} \quad , \quad z = 0^+, r \leq r_0 \quad . \tag{11}$$

The total current passing through the electrode $I$, whether directly applied to the electrode or as a response to an applied voltage input, is given by

$$I(t) = \int_0^{r_0} J_0(r,t) \cdot 2\pi r \, dr \quad . \tag{12}$$

Thus, the disk electrode system becomes solvable given (3)–(5), (11), and (12).

**Rotational Elliptic Coordinates**

Given the complexity of the set of equations, it is natural to use Fourier series methods for solving differential equations and expressing the solution with basis functions. In the cylindrical coordinates, Bessel functions can be used as the basis for the radial coordinates to solve for the primary distribution (Wiley and Webster, 1982), whereas an incorrect choice of the spherical coordinates with spherical harmonics as basis functions will lead to an dead end (Newman and Battaglia, 2018, Ch. 17).

The ideal coordinate system for the disk electrode is the rotational elliptic coordinates $(\xi, \eta, \phi)$ as shown in Figure 3, which results from rotating a two-dimensional elliptic coordinate system about the non-focal axis of the ellipse, i.e., the symmetry axis that separates the foci. Particularly, the foci are chosen to be at the edges of the electrode, so that the cylindrical coordinates $(r, \phi, z)$ are related by

$$\begin{cases} r = r_0 \cdot \sqrt{(1+\xi^2)(1-\eta^2)} \\ z = r_0 \cdot \xi \cdot \eta \\ \phi = \phi \end{cases}, \tag{13}$$

with $\eta \in [0,1]$, $\xi \in [0, +\infty)$, and $\phi \in [0, +2\pi)$.

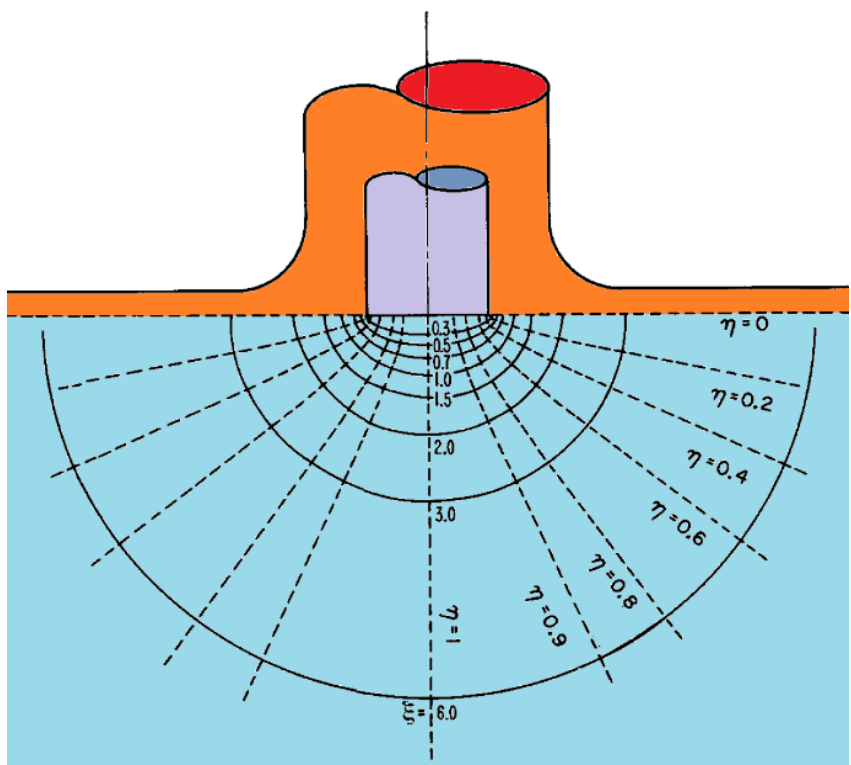

Figure 3. The disk electrode and electrolyte space in rotational elliptic coordinates. $\xi$ is a normalized distance from the "center" of the coordinate system, i.e., the disk electrode, and $\eta$ is the cosine of the "latitude", with the electrode and insulator residing on the "equatorial plane". Reprinted and adapted with permission from Newman (1966a), *J. Electrochem. Soc.*, 113(5), 501. Copyright 1966, The Electrochemical Society.

The rotational elliptic coordinates are an alternative definition of oblate spheroidal coordinates $(\mu, \nu, \phi)$ with $\xi = \sinh\mu$ and $\eta = \cos\nu$ (Newman, 1966a; Myland and Oldham, 2005): $\xi$ is analogous to the radial coordinate in a spherical coordinate system, $\eta$ is the cosine of the latitude $\nu$, $\phi$ is the longitude, the electrode and insulator are on the "equatorial plane", and the "center" of the coordinate system is not a point but the entire disk. The inverse relationships of the coordinates are

$$\begin{cases} \xi = \dfrac{\sqrt{((\hat{r}+1)^2 + \hat{z}^2)((\hat{r}-1)^2 + \hat{z}^2)} + (\hat{r}^2 + \hat{z}^2 - 1)}{2} \\ \eta = \dfrac{\sqrt{((\hat{r}+1)^2 + \hat{z}^2)((\hat{r}-1)^2 + \hat{z}^2)} - (\hat{r}^2 + \hat{z}^2 - 1)}{2} \end{cases}, \quad (14)$$

with $\hat{r} = r/r_0$ and $\hat{z} = z/r_0$ being normalized cylindrical coordinates, and the boundaries and axis of symmetry are related to their original definitions by

$$\begin{cases} z = 0^+, r \leq r_0 \\ z = 0, r > r_0 \\ r, z \to +\infty \\ r = 0, z \geq 0 \end{cases} \Leftrightarrow \begin{cases} \xi = 0^+, \eta \in [0,1] \\ \eta = 0, \xi \in (0, +\infty) \\ \xi \to +\infty, \eta \in [0,1] \\ \eta = 1, \xi \in [0, +\infty) \end{cases}. \quad (15)$$

These lines and other equal-$\xi$ lines and equal-$\eta$ lines are also shown in Figure 3.

Laplace's equation in rotational elliptic coordinates with axial symmetry becomes

$$\nabla^2 \varphi(\xi, \eta) = \frac{\partial}{\partial \xi}\left[(1+\xi^2)\frac{\partial \varphi(\xi,\eta)}{\partial \xi}\right] + \frac{\partial}{\partial \eta}\left[(1-\eta^2)\frac{\partial \varphi(\xi,\eta)}{\partial \eta}\right] = 0 \quad . \quad (16)$$

The current density right above the electrode is given by

$$J_z(r, 0^+) = -\kappa \left.\frac{\partial \varphi(r,z)}{\partial z}\right|_{z=0^+} = -\frac{\kappa}{r_0 \eta} \left.\frac{\partial \varphi(\xi,\eta)}{\partial \xi}\right|_{\xi=0^+} = J_\xi(0^+, \eta) \quad, \quad r \leq r_0 \text{ or } \eta \in [0,1] \quad, \quad (17)$$

with $\eta|_{z=0^+} = \sqrt{1 - (r/r_0)^2}$, $r \in [0, r_0]$. On the other hand, on the substrate the boundary condition becomes

$$J_z(r, 0) = -\kappa \left.\frac{\partial \varphi(r,z)}{\partial z}\right|_{z=0} = -\frac{\kappa}{r_0 \xi} \left.\frac{\partial \varphi(\xi,\eta)}{\partial \eta}\right|_{\eta=0} = J_\eta(\xi, 0) = 0 \quad, \quad r > r_0 \text{ or } \xi \geq 0 \quad, \quad (18)$$

with $\xi|_{z=0} = \sqrt{(r/r_0)^2 - 1}$, $r \in [r_0, +\infty)$.

Integration of any function $f$ on the disk surface is given by

$$\int_0^{r_0} f(r, 0) \cdot 2\pi r dr \Rightarrow 2\pi r_0^2 \int_0^1 f(0, \eta) \cdot \eta d\eta \quad, \quad (19)$$

which is used to calculate the total current or charge on the electrode–electrolyte interface.

**Basis Functions and General Form of Solution**

Using separation of variables, the potential in the electrolyte space is set to

$$\varphi(\xi,\eta) = N(\eta)M(\xi) \quad , \tag{20}$$

and Laplace's equation (16) then becomes two ordinary second order differential equations

$$\frac{d}{d\eta}\left[(1-\eta^2)\frac{dN(\eta)}{d\eta}\right] + \lambda N(\eta) = 0 \quad , \tag{21}$$

$$\frac{d}{d\xi}\left[(1+\xi^2)\frac{dM(\xi)}{d\xi}\right] - \lambda M(\xi) = 0 \quad . \tag{22}$$

Despite the ordering of the rotational elliptic coordinates having $\xi$ appear first, from here on, the equations and functions involving $\eta$ are typically placed first as they concern the system's behavior on the electrode surface and are more important. They are also relatively simpler and easier to handle.

The solutions are Legendre functions for (21) and Legendre functions with imaginary argument for (22). Let $\lambda = l(l+1)$, then

$$\begin{cases} N_l(\eta) = c_l^{NP}P_l(\eta) + c_l^{NQ}Q_l(\eta) \\ M_l(\xi) = c_l^{MP}P_l(\xi/i) + c_l^{MQ}Q_l(\xi/i) \end{cases} , \tag{23}$$

with $P_l(\eta)$ and $Q_l(\eta)$ being the $l$th order Legendre functions of the first and second kind, respectively. The boundary conditions (4) and (5) in the new coordinate system yields the following conditions

$$\begin{cases} |\varphi(\xi,\eta)| < +\infty & \Rightarrow |N_l(\eta)|, |M_l(\xi)| < +\infty \Rightarrow l \in \mathbb{N}^0, \quad c_l^{NQ} = 0 \\ -\dfrac{\kappa}{r_0\xi}\dfrac{\partial\varphi(\xi,\eta)}{\partial\eta}\bigg|_{\eta=0} = 0 & \Rightarrow \dfrac{dP_l(\eta)}{d\eta}\bigg|_{\eta=0} = 0 \quad \Rightarrow l = 2n, \ n \in \mathbb{N}^0 \\ \lim_{\xi\to+\infty}\varphi(\xi,\eta) = 0 & \Rightarrow \lim_{\xi\to+\infty} M_{2n}(\xi) = 0 \end{cases} , \tag{24}$$

which apply to all situations. For $M_l(\xi)$, the Legendre functions are evaluated on the imaginary axis. The complex coefficients $c_{2n}^{MP}$ and $c_{2n}^{MQ}$ are chosen so that $M_{2n}(\xi)$ is a real function on $\xi \in [0, +\infty)$ with the conditions in (24) satisfied and normalized so that $M_{2n}(0) = 1$. See Appendix A for details on the Legendre functions and a derivation for $M_{2n}(\xi)$, $M'_{2n}(0)$, and their properties, which were omitted in Newman's original work for brevity (Newman, 1966b, equations [14]–[15] and [19]).

Combining and renaming the coefficients, the general solution can be written in a form of summation

$$\varphi(\xi,\eta) = \sum_{n=0}^{+\infty} B_n P_{2n}(\eta) M_{2n}(\xi) \quad . \tag{25}$$

Here, $P_{2n}(\eta)M_{2n}(\xi)$ are dimensionless and normalized basis functions of the disk electrode system. The coefficients $B_n$, typically in units of Volts, will be determined for each specific voltage/current applied to the disk electrode by the equivalent form of boundary condition (11) in rotational elliptic coordinates.

The potential fields in the electrolyte are given for the first ten basis functions in Figure 4, which are displayed in the original cylindrical coordinates due to the distortion of geometry in rotational elliptic coordinates. The influence in the electrolyte is determined by $M_{2n}(\xi)$ of each basis function. Almost all non-trivial behaviors of the solutions will be limited within very close proximity—less than one times the radius—of the electrode surface, except for the zeroth solution which decays slower and extends its influence to about one order of magnitude further into the solution.

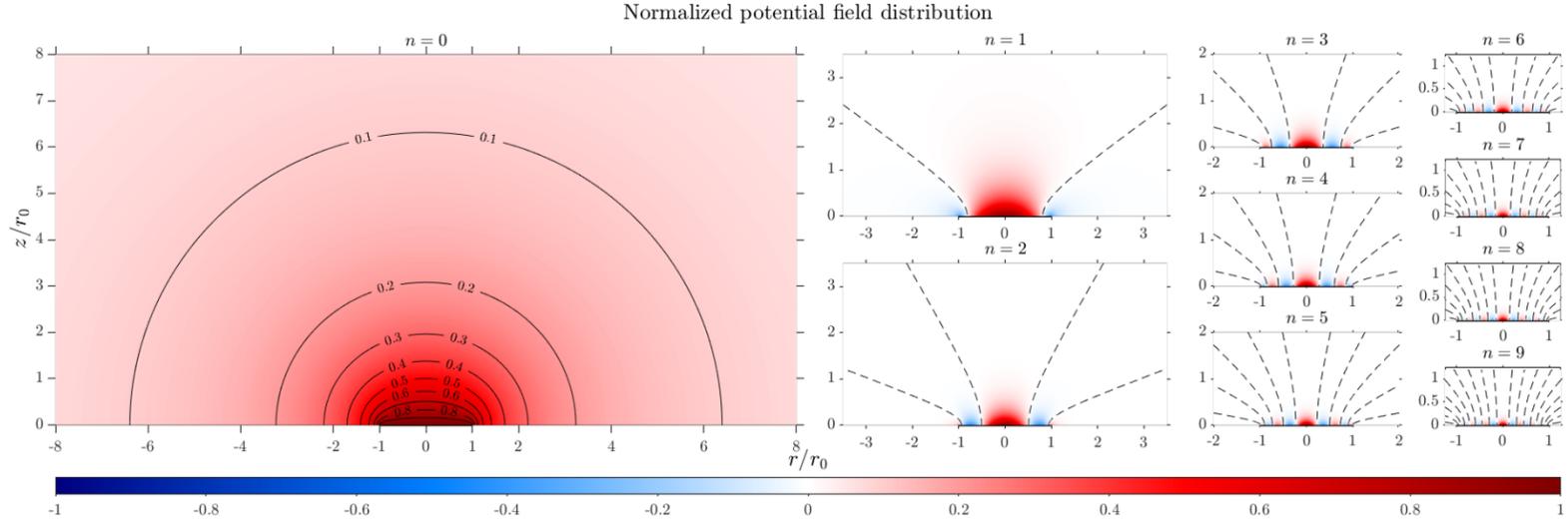

Figure 4. The dimensionless and normalized potential field distribution of the first 10 basis functions in the electrolyte space. The black line on the bottom of each panel represents the disk electrode. The field of the zeroth basis function is non-negative, non-oscillatory, and penetrates deep into the electrolyte space, while higher order basis functions are oscillatory and limited to the proximity of the electrode surface. Dashed lines indicate the zero potential.

Besides penetrating its field into the electrolyte, the zeroth basis function is critical in delivering the current. Notice that $P_0(\eta) \equiv 1$ and

$$\int_0^1 P_{2n}(\eta)\mathrm{d}\eta = \int_0^1 P_0(\eta)P_{2n}(\eta)\mathrm{d}\eta = \frac{\delta_{0n}}{4n+1} = \delta_{0n} \quad , \tag{26}$$

therefore, the total current through the electrode is only determined by the zeroth term

$$\begin{aligned} I &= 2\pi r_0^2 \int_0^1 \left( -\frac{\kappa}{r_0 \eta} \frac{\partial \varphi(\eta,\xi)}{\partial \xi} \bigg|_{\xi=0^+} \right) \eta \mathrm{d}\eta \\ &= -2\pi r_0 \kappa \left( \sum_{n=0}^{+\infty} B_n M'_{2n}(0) \int_0^1 P_{2n}(\eta)\mathrm{d}\eta \right) = 4 r_0 \kappa B_0 \quad . \end{aligned} \tag{27}$$

The normalized surface potential and current density distributions of the first 11 basis functions are shown in Figure 5. The zeroth term indicates a steady state of uniform electrolyte potential distribution above the electrode surface, i.e., primary distribution. All the other higher order terms will determine how much the actual distribution will deviate from this distribution without altering the total current. Interestingly, the current distributions of higher order solutions have higher magnitude at both the center and the periphery of the disk, while the lower order ones mainly have high current density at the periphery.

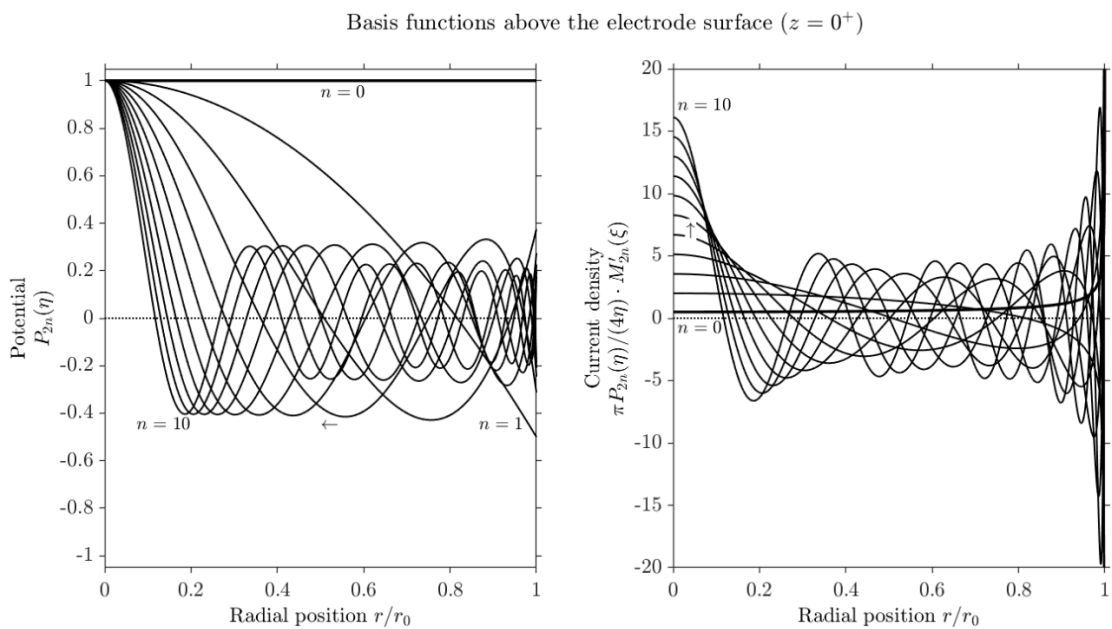

Figure 5. The normalized potential and current density distribution of the first 11 basis functions in the electrolyte above the electrode surface.

**Symbol Naming**

In the upcoming sections, the following convention is adopted for variable naming, superscripts, and subscripts, which also applies to the analysis above.

Table 1 List of symbols, subscripts, and their meaning

| | |
|---|---|
| $\varphi(r,z)$ | Potential in the electrolyte space |
| $\varphi_0(r)$ | Potential in the electrolyte immediately outside the diffusion layer: $\varphi_0(r) = \varphi(r, 0^+)$ |
| $\boldsymbol{J}(r,z)$ | Current density in the electrolyte space: $\boldsymbol{J}(r,z) = \hat{\boldsymbol{z}} \cdot J_z(r,z) + \hat{\boldsymbol{r}} \cdot J_r(r,z)$ |
| $J_0(r)$ | Current density through the interface of the electrode: $J_0(r) = J_z(r, 0^+)$ |
| $V$ | Voltage of the electrode of the interface, general case |
| $V_0$ | Amplitude of voltage step input, or primary voltage response to current step input |
| $I$ | Total current flowing across electrode-electrolyte interface, general case |
| $I_0$ | Amplitude of current step input, or primary current response to voltage step input |
| $\boldsymbol{A}_0, \boldsymbol{A}_1$ | Matrices with row and column indices starting with $(0,0)$ and $(1,1)$, respectively, with the latter being a submatrix of the former. |
| $\boldsymbol{a}_0, \boldsymbol{a}_1$ | Column vectors with the row indices starting with $0$ and $1$, respectively, with the latter being a submatrix of the former. |

Table 2 List of superscripts and their meaning

| | |
|---|---|
| None | General situation, e.g., $\varphi(r,z)$ |
| P | Primary distribution, e.g., $\varphi^P(r,z)$ |
| H | Harmonic oscillation#, e.g., $J_z(r,z,t) = J_z^H(r,z)e^{\mathrm{i}\omega t}$ |
| SS | Steady state response, e.g., $J_0^{SS}(r)$ |
| TZ | Transient response, e.g., $V^{TZ}(t)$ |
| $(i)$ | The $i$th # eigenfunction of the transient response, e.g., $B_n^{(i)}$ |

# The imaginary unit, $\mathrm{i} = \sqrt{-1}$, is given in roman font to distinguish it from the index $i$, given in italic font.

## Primary Distribution ([Newman, 1966a](#))

### Current Distribution and Electrode Resistance

This is the solution for the disk electrode without considering the surface overpotential related to the double layer and reaction currents (or infinite large double layer capacitance or reaction conductance depending on type of input). The potential and current distribution is completely determined by the ohmic resistance of the electrolyte. It is the primary distribution at $t = 0^+$ for a disk electrode applied with step input of current or voltage. The potential on the electrode is held constant at $V_0$, and the boundary condition of the electrolyte potential immediately next to the electrode is

$$\varphi_0^P(\eta) = V_0 \quad , \quad \eta \in [0,1] \quad , \tag{28}$$

indicating that

$$B_n^P = \begin{cases} V_0 \quad , & n = 0 \\ 0 \quad , & n \neq 0 \end{cases} , \tag{29}$$

which is also obvious from discussion on (27). The solution is the zeroth order basis function

$$\varphi^P(\eta, \xi) = V_0 M_0(\xi) \quad , \tag{30}$$

which can be written in several equivalent forms:

$$\frac{\varphi^P(\xi, \eta)}{V_0} = \frac{2}{\pi}\arctan(\frac{1}{\xi}) = 1 - \frac{2}{\pi}\arctan(\xi) = \frac{2}{\pi}\arcsin(\frac{1}{\sqrt{1+\xi^2}}) \quad . \tag{31}$$

The current density on the disk and total current are therefore

$$J_0^P(\eta) = J_\xi^P(0^+, \eta) = \frac{2}{\pi} \frac{\kappa V_0}{r_0 \eta} \quad , \tag{32}$$

$$I_0 = 4\kappa r_0 V_0 \quad . \tag{33}$$

The effective series resistance of the electrolyte, also termed access resistance, is

$$R_S = \frac{V_0}{I_0} = \frac{1}{4\kappa r_0} \quad . \tag{34}$$

The surface current density, as shown in Figure 6, can also be given in the cylindrical coordinates as

$$J_0^P(r) = \frac{2}{\pi} \frac{\kappa V_0}{\sqrt{r_0^2 - r^2}} = \frac{\overline{J_0}}{2\sqrt{1 - (r/r_0)^2}} \quad , \tag{35}$$

with the average current density given as

$$\overline{J_0} = \frac{I_0}{\pi r_0^2} = \frac{4\kappa V_0}{\pi r_0} \quad . \tag{36}$$

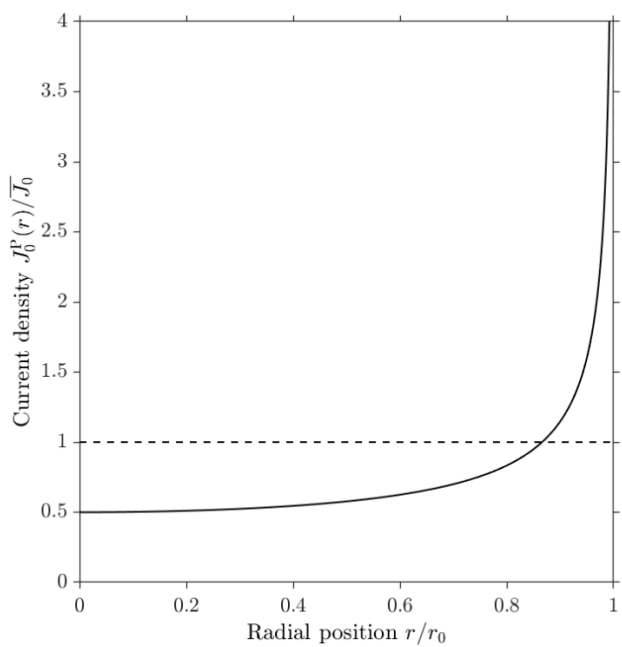

Figure 6. Normalized primary current distribution. The current density at the center is only half the average.

**Normalization Factors and Dimensionless Variables**

Besides the obvious choice of electrode radius $r_0$ for length normalization, two characteristic quantities are used for normalization in the later calculations of this review. The average electrolyte conductance from the electrode surface ($\xi = 0^+$) to ground ($\xi \to +\infty$) of the primary distribution is

$$\bar{g}_S = \frac{G_S}{\pi r_0^2} = \frac{1}{\pi r_0^2 R_S} = \frac{4}{\pi}\frac{\kappa}{r_0} \quad . \tag{37}$$

The characteristic time constant is

$$\tau = R_S C_{DL} = \frac{\pi}{4}\frac{\gamma r_0}{\kappa} \quad , \tag{38}$$

with the total double layer capacitance at steady state given as

$$C_{DL} = \gamma \pi r_0^2 \quad . \tag{39}$$

The characteristic time constant would have typical numbers in the range of microseconds to milliseconds depending on the size and material of the electrode, as well as the electrolyte. Typically values of the double layer capacitance $\gamma$ of metal electrodes are in the range between 10 and 40 μF/cm², physiological salines have conductivity $\kappa$ on the order of 0.1~1 S·m⁻¹, and electrode size $r_0$ is in the range between tens of micrometers and several millimeters. As will be seen in the , strictly speaking the time constant is frequency dependent as both the resistance and capacitance are. Also, local charging and discharging rates differ from the global process, giving rise to local time constants. The validity of the overall time constant is discussed by Oldham (2004) and $\tau$ represents the average local time constant weighted by both area and current density.

With these normalization quantities, additional dimensionless quantities are introduced. The dimensionless frequency and time are

$$\Omega = \tau\omega = \frac{\pi}{4}\frac{\gamma r_0}{\kappa}\omega \quad , \tag{40}$$

$$\theta = \frac{t}{\tau} = \frac{4\kappa}{\pi r_0 \gamma}t \quad , \tag{41}$$

and the dimensionless charge transfer conductance is

$$G = \frac{g_{CT}}{\bar{g}_S} = (\alpha_a + \alpha_c)\frac{\pi r_0 j_0 nF}{4\kappa RT} \quad , \tag{42}$$

which is similar to $J$, the dimensionless exchange current density defined by Newman (1996b, 1970b) and Nisancioğlu (1973a,b).

In practice, typical biomedical electrodes have much smaller Faradaic reactions conductance compared to the electrolyte conductance. In the Randles model, the parallel resistance of the Faradaic reactions extracted from impedance spectroscopy is typically in the megaohm to gigaohm range, compared to kiloohm values of the series resistance of the electrolyte. This gives several orders of magnitudes difference between the two quantities and hence $G$ is typically very small, i.e., ($G < 10^{-3} \ll 1$). Or in terms of conductance per unit surface area, platinum, for example, has $g_{CT}^{Pt} \approx 34$ μS·cm$^{-2}$ (Richardot and McAdams, 2002), whereas $\bar{g}_S \approx 0.01 \sim 1$ S·cm$^{-2}$ with the typical electrode and electrolyte parameters given above. However, calculations for $G = 1$ and larger values are still performed to identify the general trends.

## Frequency Dispersion ([Newman 1970a](#))

With AC input on the electrode, the current can pass from the electrode to the electrolyte by either capacitive charging of the double layer or charge transfer via Faradaic reactions. The overall equivalent circuit can be modeled by a capacitive interface, which is often characterized as a constant phase element in spectroscopy, and the electrolyte resistor in series. In this model, there is no explicit model element for the Faradaic conductance, but its effect is accounted for in the frequency dispersion of the double layer capacitance and series resistance of the electrolyte.

The electrode potential is given as

$$V(t) = V^{\mathrm{H}} \mathrm{e}^{\mathrm{i}\omega t} \quad , \tag{43}$$

and the potential in the electrolyte is normalized with respect to the electrode potential

$$\varphi(r, z, t) = \varphi^{\mathrm{H}}(r, z)\mathrm{e}^{\mathrm{i}\omega t} = V^{\mathrm{H}} U^{\mathrm{H}}(r, z)\mathrm{e}^{\mathrm{i}\omega t} \quad . \tag{44}$$

With these quantities, the boundary condition (11) on the electrode therefore becomes

$$\left.\frac{\partial U^{\mathrm{H}}(\xi, \eta)}{\partial \xi}\right|_{\xi=0+} = -\frac{4(\mathrm{i}\Omega + G)}{\pi}\eta\left(1 - U_0^{\mathrm{H}}(\eta)\right) \quad . \tag{45}$$

Obviously, the normalized potential distribution $U^{\mathrm{H}}(\eta, \xi)$ also satisfy the Laplace's equation and the same boundary conditions (4) as $\varphi(\eta, \xi)$. Therefore

$$U^{\mathrm{H}}(\xi, \eta) = \sum_{n=0}^{+\infty} B_n^{\mathrm{H}} P_{2n}(\eta) M_{2n}(\xi) \quad , \tag{46}$$

with the coefficients $B_n^{\mathrm{H}}(\Omega)$ being functions of the input frequency. Therefore, (45) is rewritten as

$$\sum_{n=0}^{+\infty} B_n^{\mathrm{H}} M'_{2n}(0) P_{2n}(\eta) = -\frac{4(\mathrm{i}\Omega + G)}{\pi}\eta\left(1 - \sum_{n=0}^{+\infty} B_n^{\mathrm{H}} P_{2n}(\eta)\right) \quad . \tag{47}$$

To determine their value, especially $B_0^{\mathrm{H}}$, (47) is multiplied by $P_{2m}(\eta)$ and integrated with respect to $\eta$ over 0 to 1. Utilizing the orthogonality of Legendre polynomials, this results in an infinite set of equations for $B_n^{\mathrm{H}}$ ([Newman 1970a](#), Equation [14])

$$\sum_{n=0}^{+\infty} B_n^{\mathrm{H}} M'_{2n}(0)\frac{\delta_{mn}}{4m+1} = -\frac{4(\mathrm{i}\Omega + G)}{\pi}\int_0^1 \eta\left(1 - \sum_{n=0}^{+\infty} B_n^{\mathrm{H}} P_{2n}(\eta)\right) P_{2m}(\eta)\mathrm{d}\eta \quad , \tag{48}$$

which can be written in matrix format

$$\left((G + \mathrm{i}\Omega)\mathbf{A}_0 + \mathbf{M}_0\right)\mathbf{b}_0^{\mathrm{H}} = (G + \mathrm{i}\Omega)\mathbf{a}_0 \quad , \tag{49}$$

where $\mathbf{b}_0^{\mathrm{H}} = \left[B_0^{\mathrm{H}}, B_1^{\mathrm{H}}, \cdots, B_n^{H}, \cdots\right]^{\mathrm{T}}$, $\mathbf{M}_0$ is a diagonal matrix

$$\boldsymbol{M_0} = \text{diag}\left(\frac{-\pi M'_{2m}(0)}{4(4m+1)}\right) \quad , \quad m \in \mathbb{N}^0 \quad , \tag{50}$$

and $\boldsymbol{A_0}$ and $\boldsymbol{a_0}$ are a matrix and a vector defined as

$$\begin{aligned} \boldsymbol{A_0} &= [a_{m,n}]_{+\infty \times +\infty} \quad , \quad m,n \in \mathbb{N}^0 \\ \boldsymbol{a_0} &= [a_{m,0}]_{+\infty \times 1} \quad , \quad m \in \mathbb{N}^0 \end{aligned} \tag{51}$$

with

$$\begin{aligned} a_{m,n} = a_{n,m} &= \int_0^1 \eta P_{2m}(\eta) P_{2n}(\eta) \, d\eta \quad , \quad m,n \in \mathbb{N}^0 \\ a_{0,n} = a_{n,0} &= -\frac{P_{2n}(0)}{2(2n-1)(n+1)} \quad , \quad n \in \mathbb{N}^0 \end{aligned} \tag{52}$$

See Tables B2 and B3 for the numeric values of these matrices.

Solving the coefficients of the basis functions gives

$$\boldsymbol{b_0^H} = \left((G + i\Omega)\boldsymbol{A_0} + \boldsymbol{M_0}\right)^{-1}(G + i\Omega)\boldsymbol{a_0} \quad , \tag{53}$$

which, for $G + i\Omega \neq 0$, is

$$\boldsymbol{b_0^H} = \left(\boldsymbol{A_0} + \frac{\boldsymbol{M_0}}{G + i\Omega}\right)^{-1}\boldsymbol{a_0} \quad . \tag{54}$$

The solution is practical only with the matrix indices truncated to a finite number $n_{\max}$ of rows and columns. See the [Numeric Accuracy](#) section for discussion on the accuracy of the solution.

With $B_0^H$ calculated, the AC current through the electrode is given by (27) as

$$I^H = 4r_0 \kappa B_0^H V^H \quad . \tag{55}$$

The complex impedance of the electrode-electrolyte system is therefore

$$Z^H = \frac{V^H}{I^H} = \frac{1}{4r_0 \kappa B_0^H} = R_S^H + \frac{1}{i\omega C_{DL}^H} \tag{56}$$

with

$$\frac{R_S^H}{R_S} = \frac{\Re(B_0^H)}{|B_0^H|^2} \quad , \quad \frac{C_{DL}}{C_{DL}^H} = \Omega \frac{\Im(B_0^H)}{|B_0^H|^2} \quad , \tag{57}$$

from comparing (56) with (34) and (39). Hence, the frequency dispersion is obtained.

Figure 7 shows the normalized impedance spectrum of the disk electrode for a few selected values of $G$. The spectrum can be very closely fitted with a three-element circuit consisting of the electrolyte resistance in series with a parallel connection of the charge transfer resistance and the double layer either represented by a capacitor or a constant phase element (CPE). The frequency dispersion of the resistive and capacitive impedance is also shown in Figure 8. The capacitive impedance appears to peak at $\Omega = G$ and exhibits a CPE behavior that deviates from the −1 log-log slope for higher frequency (Huang et al., 2007a, b), as shown in Figure 9.

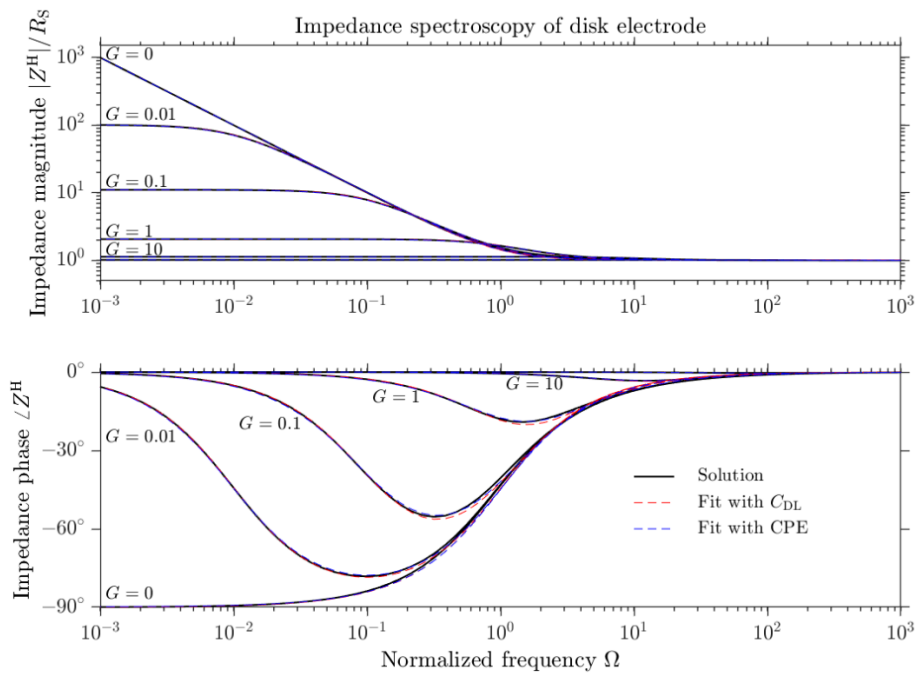

Figure 7. Impedance spectrum of the disk electrode for different Faradaic conductance values. Dashed line indicates best fits with two equivalent circuit models.

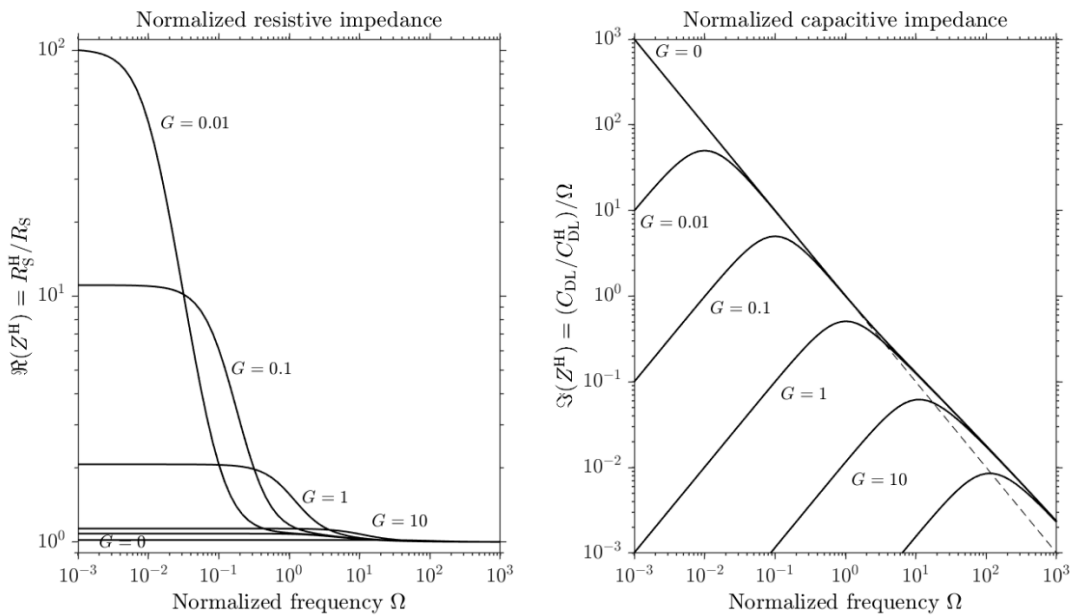

Figure 8. Equivalent resistive and capacitive impedance of the disk electrode for different Faradaic conductance values. The dashed line in the right panel has a slope of −1.

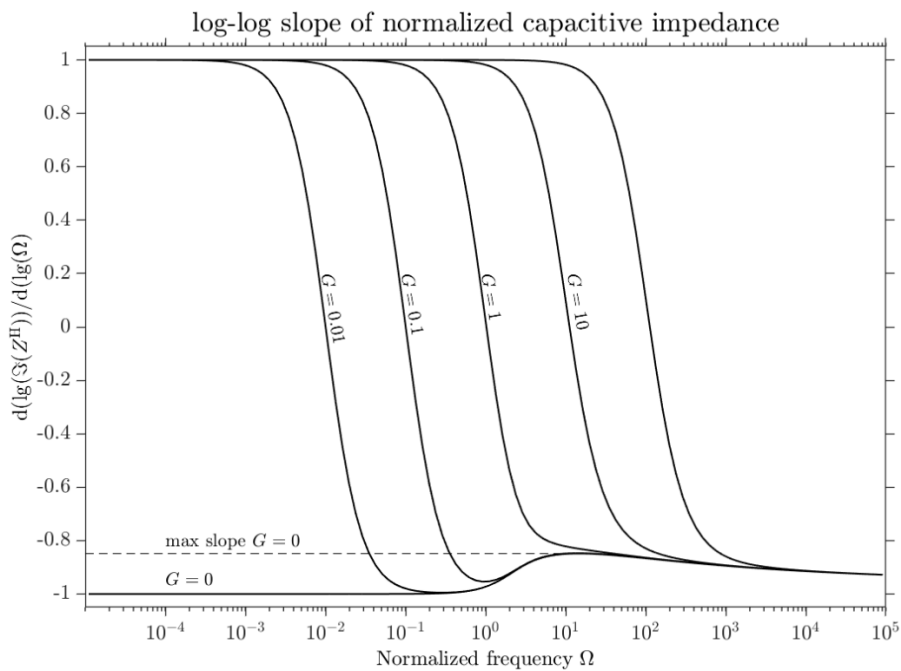

Figure 9. The slope of the capacitive impedance of the disk electrode for different Faradaic conductance values, showing the CPE behavior at high frequency.

With $\boldsymbol{b}_0^H$, the potential and current density distribution are calculated (Figure 10). For $\Omega \to 0$, the distributions are the same as the steady state response for voltage step input. The distributions converge on the primary distributions either for $\Omega \to +\infty$, regardless of $G$, or for $G \to +\infty$, regardless of $\Omega$.

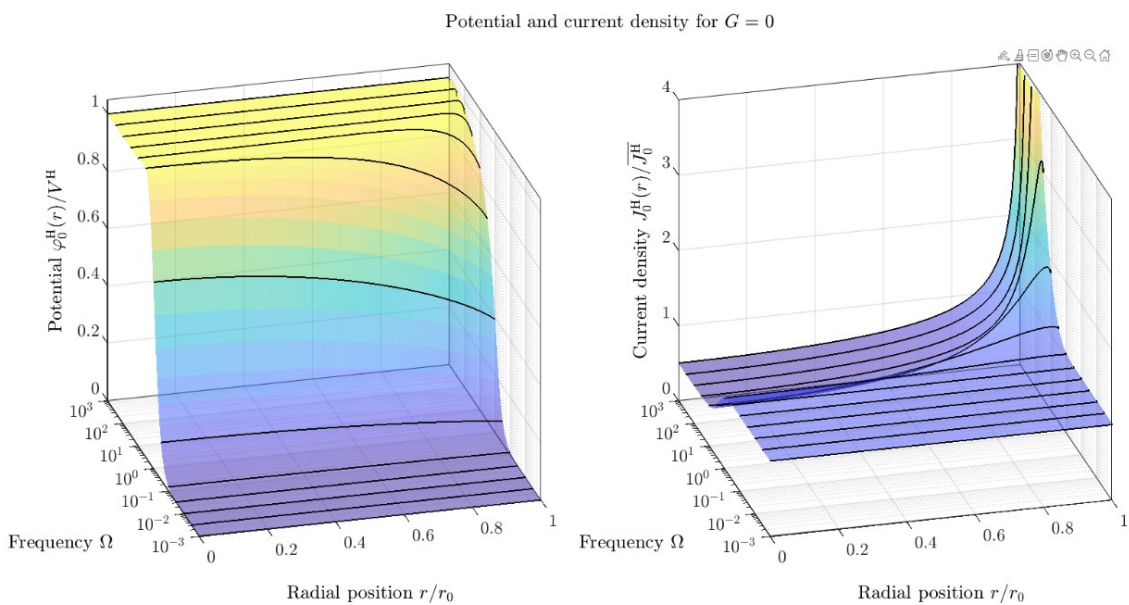

Figure 10. The potential and current density distributions above the electrode for a wide range of normalized frequencies and several charge transfer reaction conductances. Figure continues on next page.

Potential and current density for $G = 0.1$

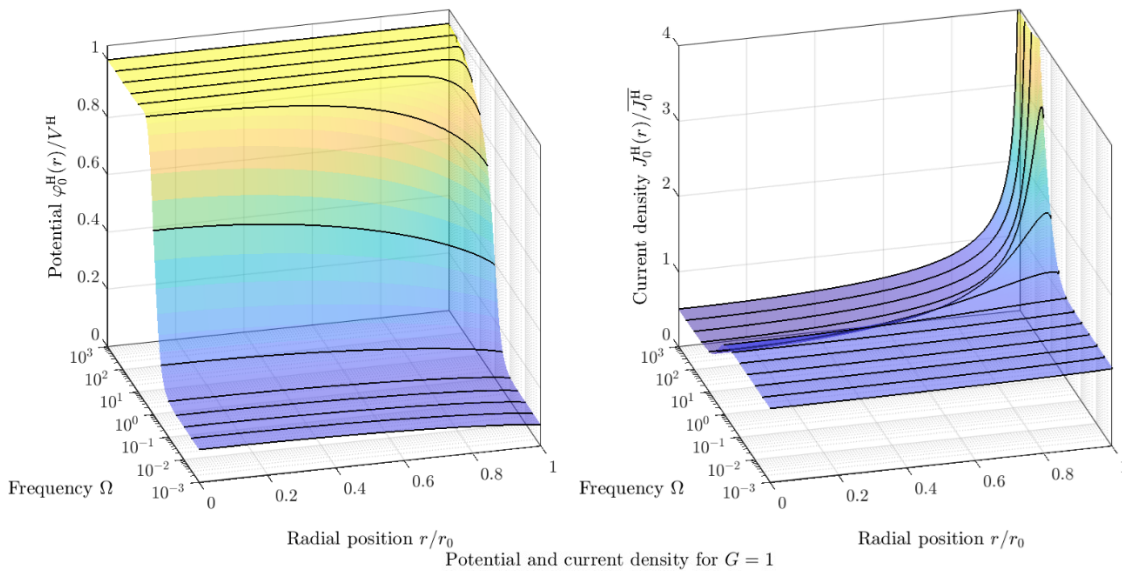

Potential and current density for $G = 1$

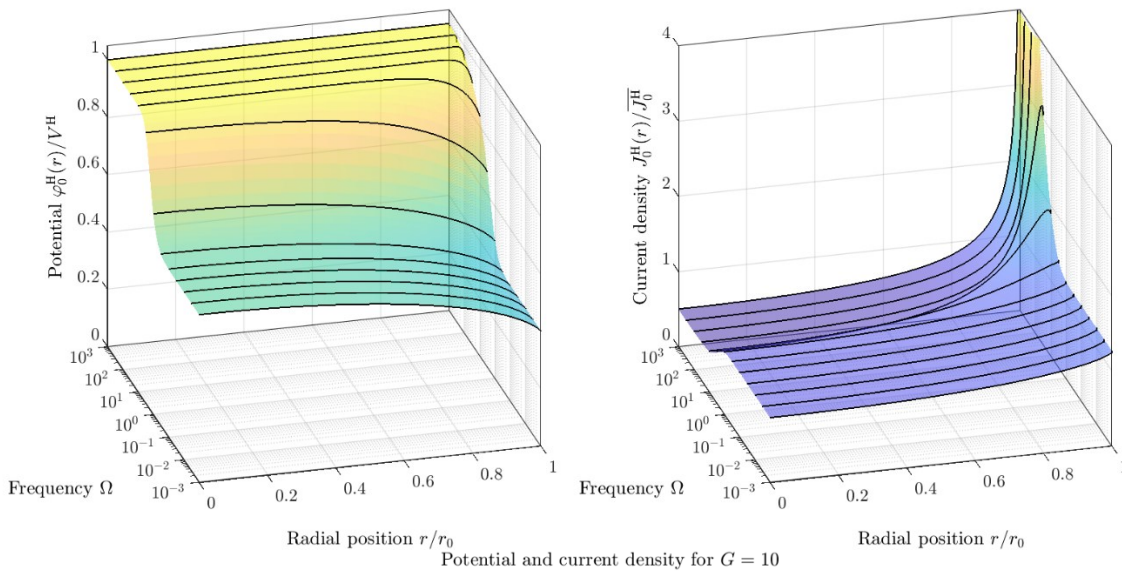

Potential and current density for $G = 10$

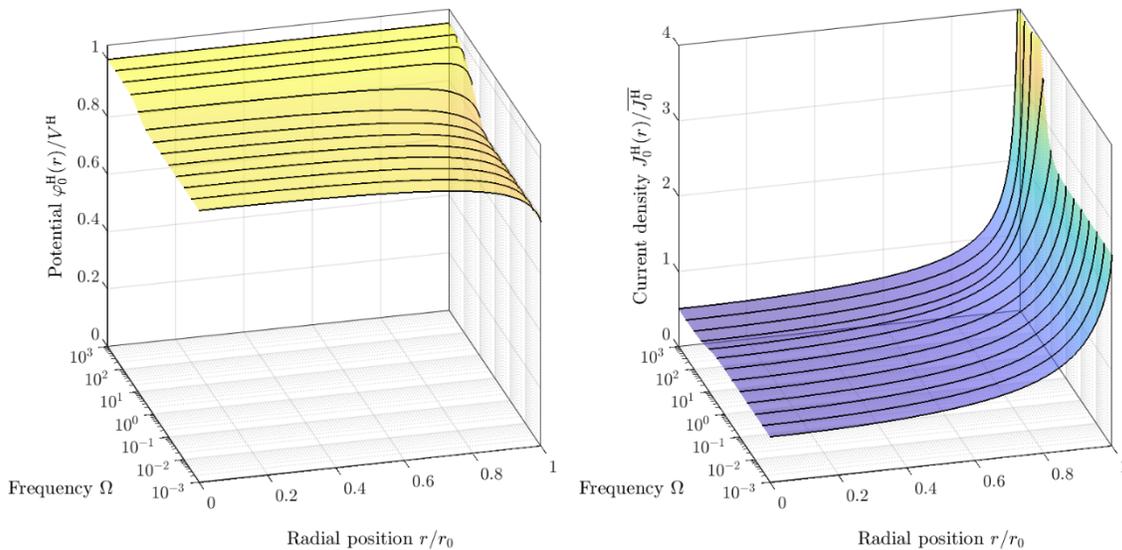

## Transient Response to Current Step Input ([Nisancioğlu and Newman, 1973a](#))

The transient response to current step input

$$I(t) = I_0 \cdot u(t) \quad , \tag{58}$$

applied to the electrode has the same boundary condition (11) as the sinusoidal situation.

**Decomposition of the Solution**

The solution can be given as a decomposition of a steady-state response and transient contribution,

$$\varphi(r,z,t) = \varphi^{SS}(r,z) \cdot u(t) - \varphi^{TZ}(r,z,t) \quad , \tag{59}$$

in which the former includes the contribution of the input current that stays constant ($I(t) \equiv I_0 \equiv I^{SS}$), and the later represents the shift from the initial condition to the steady-state solution and does not contribute to the electrode current. The initial condition immediately after the input's onset ($t = 0^+$) is easily shown to be identical to the primary distribution. Therefore, the current and potential values $I_0$ and $V_0$ for the primary distribution are used as normalization factors for the solution afterwards. This gives the convenience to set the coefficient of the zeroth term in the steady-state solution

$$\frac{\varphi^{SS}(\eta,\xi)}{V_0} = \sum_{n=0}^{+\infty} B_n^{SS} P_{2n}(\eta) M_{2n}(\xi) \tag{60}$$

to unity ($B_0^{SS} = 1$), as the total current input is always the same during the transition. The voltage on the electrode as well as other variables could also be decomposed in the same manner

$$V(t) = V^{SS} \cdot u(t) - V^{TZ}(t) \quad . \tag{61}$$

**Steady State Response**

Analyzing the general boundary condition (11) for the steady state gives

$$\left.\frac{\partial \varphi^{SS}(\eta,\xi)}{\partial \xi}\right|_{\xi=0+} = -\frac{4G}{\pi}\eta\left(V^{SS} - \varphi_0^{SS}(\eta)\right) \quad . \tag{62}$$

Utilizing the same method for the frequency dispersion problem, the equation is multiplied by $P_{2m}(\eta)$ and integrated with respect of $\eta$ over 0 to 1 after substituting (60) into (62)

$$a_{m,0}G\frac{V^{SS}}{V_0} = \sum_{n=0}^{+\infty} G a_{m,n} B_n^{SS} - \frac{\pi M'_{2m}(0)B_m^{SS}}{4(4m+1)} \quad , \quad m \in \mathbb{N}^0 \quad . \tag{63}$$

where $a_{m,n}$ are defined in (52). This yields Equations [10] and [11] of Nisancioğlu and Newman ([1973a](#)). Specifically, for $m = 0$, with $B_0^{SS} = 1$ and $a_{0,0} = 1/2$, condition (63) becomes

$$G \frac{V^{SS}}{V_0} = 2\left(G \sum_{n=0}^{+\infty} B_n^{SS} a_{0,n} - \frac{\pi M_0'(0) B_0^{SS}}{4}\right) = G + 1 + 2G \sum_{n=1}^{+\infty} a_{0,n} B_n^{SS} \quad , \tag{64}$$

and for $m \in \mathbb{N}^+$, (64) is substituted into the left side of (63), yielding

$$\sum_{n=1}^{+\infty} \left[G(a_{m,n} - 2a_{m,0} a_{0,n}) - \frac{\pi M_{2m}'(0)\delta_{mn}}{4(4m+1)}\right] B_n^{SS} = a_{m,0} \quad . \tag{65}$$

Let $\boldsymbol{b}_1^{SS} = [B_1^{SS}, B_2^{SS}, \cdots, B_n^{SS}, \cdots]^T$, the group of equations in (65) can be written in matrix form

$$\left(G(\boldsymbol{A}_1 - 2\boldsymbol{a}_1 \boldsymbol{a}_1^T) + \boldsymbol{M}_1\right)\boldsymbol{b}_1^{SS} = \boldsymbol{a}_1 \quad , \tag{66}$$

where $\boldsymbol{A}_1$, $\boldsymbol{a}_1$, and $\boldsymbol{M}_1$ are matrixes are sub-matrices of $\boldsymbol{A}_0$, $\boldsymbol{a}_0$, and $\boldsymbol{M}_0$, respectively, with row and column indices starting from 1 instead 0. See Tables B2 and B3 for the numeric values of these matrices.

Solving the coefficients of the basis functions of the Laplace's equation with the matrix indices truncated to a finite number $n_{\max}$ of rows and columns gives

$$\boldsymbol{b}_1^{SS} = \left(G(\boldsymbol{A}_1 - 2\boldsymbol{a}_1 \boldsymbol{a}_1^T) + \boldsymbol{M}_1\right)^{-1} \boldsymbol{a}_1 \quad . \tag{67}$$

Alternatively, (63) and (64) can be directly written as

$$(G\boldsymbol{A}_0 + \boldsymbol{M}_0)\boldsymbol{b}_0^{SS} = G \frac{V^{SS}}{V_0} \boldsymbol{a}_0 \quad , \tag{68}$$

and

$$\frac{V^{SS}}{V_0} = G^{-1} + 2\boldsymbol{a}_0^T \boldsymbol{b}_0^{SS} \quad , \tag{69}$$

where $\boldsymbol{b}_0^{SS} = [B_0^{SS}, B_1^{SS}, \cdots, B_n^{SS}, \cdots]^T$. Combining (68) and (69) yields

$$\left(G(\boldsymbol{A}_0 - 2\boldsymbol{a}_0 \boldsymbol{a}_0^T) + \boldsymbol{M}_0\right)\boldsymbol{b}_0^{SS} = \boldsymbol{a}_0 \quad , \tag{70}$$

and therefore

$$\boldsymbol{b}_0^{SS} = \left(G(\boldsymbol{A}_0 - 2\boldsymbol{a}_0 \boldsymbol{a}_0^T) + \boldsymbol{M}_0\right)^{-1} \boldsymbol{a}_0 \quad . \tag{71}$$

With $\boldsymbol{b}_0^{SS}$ or $\boldsymbol{b}_1^{SS}$ solved (see Table B4 for numeric values), the steady state solution of the electric field is obtained via (60). The potential and current density distributions in the electrolyte are shown in Figure 11. For $G \to +\infty$, the steady-state distributions converge to the primary distributions, whereas for $G \to 0$, the current density is uniform. The average potential in the electrolyte above the electrode is

$$\begin{aligned}\overline{\frac{\varphi_0^{SS}}{V_0}} &= \left(2\pi r_0^2 \int_0^1 \sum_{n=0}^{+\infty} B_n^{SS} P_{2n}(\eta)\, \eta\, d\eta\right)/(\pi r_0^2) = 2\boldsymbol{a}_0^T \boldsymbol{b}_0^{SS} \\ &= 2\boldsymbol{a}_0^T \left(G(\boldsymbol{A}_0 - 2\boldsymbol{a}_0 \boldsymbol{a}_0^T) + \boldsymbol{M}_0\right)^{-1} \boldsymbol{a}_0 \quad .\end{aligned} \tag{72}$$

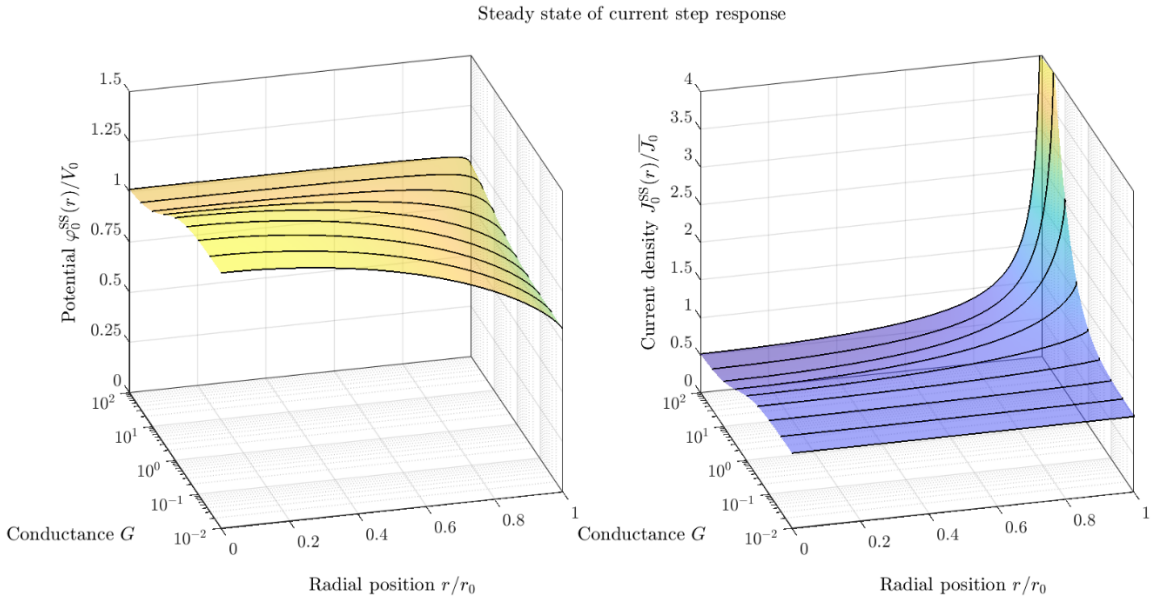

Figure 11. The steady state potential and current density distribution on the surface of the electrode for different Faradaic reaction conductance.

The steady state electrode voltage is given according to (69) as

$$\frac{V^{SS}}{V_0} = G^{-1} + 2\boldsymbol{a}_0^T \boldsymbol{b}_0^{SS} = G^{-1} + \frac{\overline{\varphi_0^{SS}}}{V_0} \quad , \quad G > 0 \quad , \tag{73}$$

and is shown in Figure 12 as a function of the Faradaic conductance. The average overpotential is inversely proportional to the Faradaic conductance $\overline{V_{OP}^{SS}}/V_0 = G^{-1}$. Obviously, a steady state solution does not exist for $G \to 0$, which is discussed later.

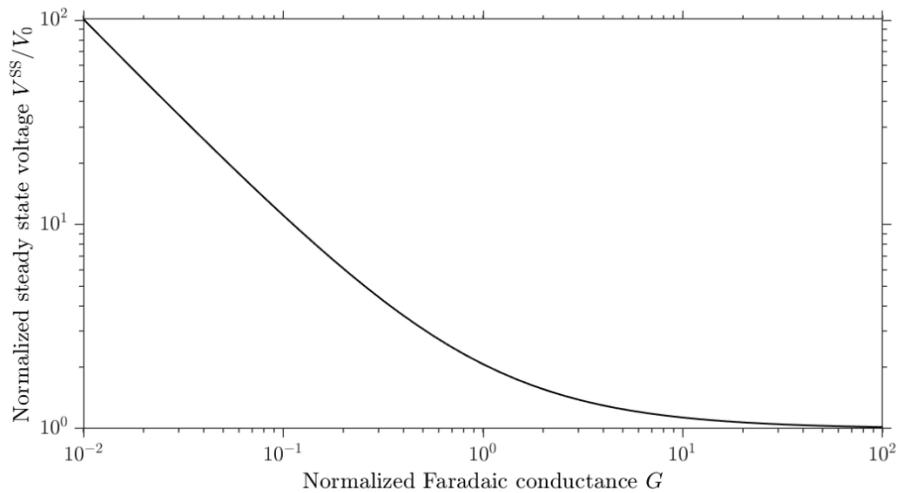

Figure 12. The steady state voltage of the electrode for different Faradaic reaction conductance.

**Eigenfunctions of the Transient Response**

The transient response does not contribute to the current input, and only redistributes the potential and current density throughout the entire space from their initial state to the steady state. Because the boundary condition (11) is a first order differential equation in terms of time, the solution could be assumed to be exponential decays of certain spatial eigenfunctions using the method of variable separation

$$\frac{\varphi^{\text{TZ}}(r,z,t)}{V_0} = \left[\sum_{i=0}^{+\infty} C^{(i)} e^{-\frac{t}{\tau^{(i)}}} U^{(i)}(r,z)\right] \cdot u(t) \quad, \tag{74}$$

where $\tau^{(i)}$ is the time constant of the decay, $C^{(i)}$ is the coefficient for the decay, and $U^{(i)}(r,z)$ is the spatial distribution of the field associated with the $i$th decay. It should be noted that, unlike Nisancioğlu and Newman ([1973a](#)), here the order of the eigenfunctions is not directly assumed to start from one and includes the zeroth order eigenfunction. It will be shown that the zeroth order eigenvalue and eigenfunction equal zero, thus being consistent with physical intuition.

Switching to rotational elliptical coordination and introducing the dimensionless eigenvalue

$$\Lambda^{(i)} = \frac{\tau}{\tau^{(i)}} - G \quad\Leftrightarrow\quad \tau^{(i)} = \frac{\tau}{\Lambda^{(i)} + G} \quad, \tag{75}$$

the solution to the potential in the electrolyte is

$$\frac{\varphi^{\text{TZ}}(\eta,\xi,\theta)}{V_0} = \left[\sum_{i=0}^{+\infty} C^{(i)} e^{-\theta(\Lambda^{(i)}+G)} U^{(i)}(\eta,\xi)\right] \cdot u(\theta) \quad, \tag{76}$$

with $U^{(i)}(\eta,\xi)$ given as

$$U^{(i)}(\eta,\xi) = \sum_{n=1}^{+\infty} B_n^{(i)} P_{2n}(\eta) M_{2n}(\xi) \quad. \tag{77}$$

The transient part of the voltage on the electrode can be given as

$$\frac{V^{\text{TZ}}(\theta)}{V_0} = \left[\sum_{i=0}^{+\infty} C^{(i)} e^{-\theta(\Lambda^{(i)}+G)}\right] \cdot u(\theta) \quad, \tag{78}$$

with equal coefficients $C^{(i)}$. This is possible as the spatial distributions $U^{(i)}(\eta,\xi)$ could be scaled by their coefficients $B_n^{(i)}$.

For each transient decay, the general boundary condition (11) becomes

$$\left.\frac{\partial U^{(i)}(\eta,\xi)}{\partial \xi}\right|_{\xi=0+} = \frac{4\eta}{\pi} \Lambda^{(i)} \left(1 - U_0^{(i)}(\eta)\right) \quad, \tag{79}$$

with $B_0^{(i)} = 0$ from the analysis. The specific boundary condition (79) then becomes

$$\sum_{n=1}^{+\infty} B_n^{(i)} P_{2n}(\eta) M'_{2n}(0) = \frac{4\eta}{\pi} \Lambda^{(i)} \left(1 - \sum_{n=1}^{+\infty} B_n^{(i)} P_{2n}(\eta)\right) \quad . \tag{80}$$

Let $\boldsymbol{b}_1^{(i)} = \left[B_1^{(i)}, B_2^{(i)}, \cdots, B_n^{(i)}, \cdots\right]^{\mathrm{T}}$, a trivial solution exists and is denoted as the zeroth order eigenfunction:

$$\begin{aligned} \Lambda^{(0)} &= 0 \\ \boldsymbol{b}_1^{(0)} &= \boldsymbol{0}_1 \end{aligned} \quad . \tag{81}$$

For $\Lambda^{(i)} \neq 0$, (80) is multiplied by $P_{2m}(\eta)$ and again integrated with respect to $\eta$ over 0 to 1, yielding equations [26] and [27] of Nisancioğlu and Newman ([1973a](#)). Specifically,

$$\sum_{n=1}^{+\infty} a_{0,n} B_n^{(i)} = a_{0,0} = \frac{1}{2} \quad , \quad m = 0 \quad , \tag{82}$$

$$\sum_{n=1}^{+\infty} \left[\Lambda^{(i)} a_{m,n} + \frac{\pi M'_{2m}(0)\delta_{mn}}{4(4m+1)}\right] B_n^{(i)} = \Lambda^{(i)} a_{m,0} \quad , \quad m \in \mathbb{N}^+ \quad , \tag{83}$$

which, written in matrix format are

$$2\boldsymbol{a}_1^{\mathrm{T}} \boldsymbol{b}_1^{(i)} = 1 \quad , \tag{84}$$

$$\left(\Lambda^{(i)} \boldsymbol{A}_1 - \boldsymbol{M}_1\right) \boldsymbol{b}_1^{(i)} = \Lambda^{(i)} \boldsymbol{a}_1 \quad , \tag{85}$$

respectively. Left-multiplying both side of (84) with $\boldsymbol{a}_1$ yields

$$2(\boldsymbol{a}_1 \boldsymbol{a}_1^{\mathrm{T}}) \boldsymbol{b}_1^{(i)} = \boldsymbol{a}_1 \quad , \tag{86}$$

and substituting (86) into (85) gives

$$\left(\boldsymbol{M}_1 - \Lambda^{(i)}\left(\boldsymbol{A}_1 - 2\boldsymbol{a}_1 \boldsymbol{a}_1^{\mathrm{T}}\right)\right) \boldsymbol{b}_1^{(i)} = 0 \quad . \tag{87}$$

Therefore, $\Lambda^{(i)}$ are the eigenvalues of the positive-definite matrix $\boldsymbol{M}_1 \left(\boldsymbol{A}_1 - 2\boldsymbol{a}_1 \boldsymbol{a}_1^{\mathrm{T}}\right)^{-1}$. For numeric calculation, all matrixes are truncated to $n_{\max}$ of rows and columns, and the first $n_{\max}$ eigenvalues could be obtained in ascending order (descending value for $\tau^{(i)}$). The corresponding coefficients $\boldsymbol{b}_1^{(i)}$ can then be obtained from (85):

$$\boldsymbol{b}_1^{(i)} = \left(\boldsymbol{A}_1 - \frac{\boldsymbol{M}_1}{\Lambda^{(i)}}\right)^{-1} \boldsymbol{a}_1 \quad . \tag{88}$$

See Table B5 for the numeric values of $\Lambda^{(i)}$ and $\boldsymbol{b}_1^{(i)}$ obtained with $n_{\max} = 1024$.

As can be observed from (79)–(87), the eigenvalues and spatial distribution of the transient response's eigenfunctions are independent of the presence or magnitude of the Faradaic reaction. The normalized potential and current density distributions of the eigenfunctions with order greater than zero are shown in Figure 13. They correspond to the charge redistribution of the double layer capacitance through the electrolyte (Figure 14). The current density on the electrode is proportional to $V^{TZ}(\theta) - \varphi_0^{TZ}(\eta, \theta)$ or $1 - U_0^{(i)}(\eta)$ according to (79). The current density of one eigenfunction is orthogonal to the potential of another:

$$\int_0^1 U_0^{(i)}(\eta)\left(1 - U_0^{(j)}(\eta)\right)\eta d\eta = \frac{\delta_{ij}}{\Lambda^{(i)}} \sum_{n=1}^{+\infty} \frac{\pi M'_{2n}(0)}{4(4n+1)} \left(B_n^{(i)}\right)^2 = -\frac{\delta_{ij}}{\Lambda^{(i)}} \left(\boldsymbol{b}_1^{(i)}\right)^T \boldsymbol{M}_1 \boldsymbol{b}_1^{(i)} \quad , \quad (89)$$

which is utilized to obtain the coefficient $C^{(i)}$ for the exponential decay. Despite the non-existence of the zeroth order eigenfunction $U^{(0)}$ in the electrolyte, its counterpart $C^{(0)}$ on the electrode (78) exists, and represents the local charge redistribution via Faradaic reaction with time constant $\tau^{(0)} = \tau/G$ (Figure 14).

**Transient Response**

The initial condition after input onset ($t = 0^+$) is the primary distribution. For the potential, (59) could be evaluated at $t = 0^+$ on the electrode surface ($z = \xi = 0^+$) giving

$$\begin{aligned}\frac{\varphi(r, 0^+, 0^+)}{V_0} &= 1 = \left[\frac{\varphi^{SS}(r, 0^+)}{V_0} - \frac{\varphi^{TZ}(r, 0^+, 0^+)}{V_0}\right] \cdot u(0^+) \\ &= \frac{\varphi_0^{SS}(\eta)}{V_0} - \sum_{j=1}^{+\infty} C^{(j)} U_0^{(j)}(\eta) \quad ,\end{aligned} \quad (90)$$

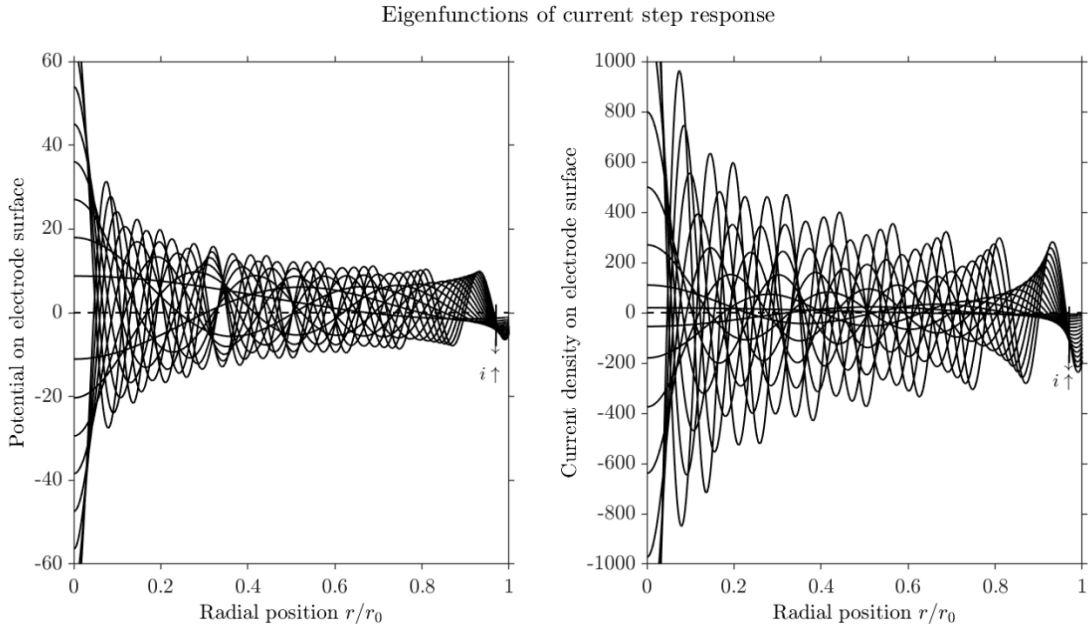

Figure 13. The first 16 normalized and dimensionless "potential" and "current density" distribution on the surface of the electrode of the eigenfunctions of the transient response to current step input.

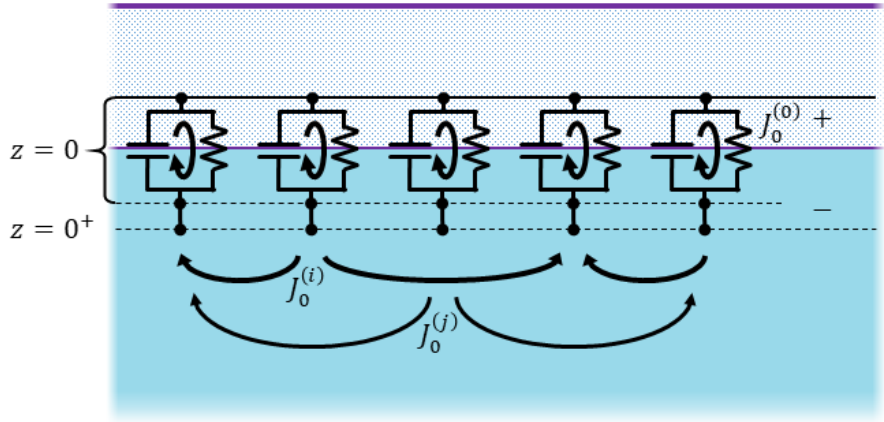

Figure 14. Charge redistribution of the double layer via different pathways during the transient. The zeroth eigenfunction distribute charge through local Faradaic reaction, while higher order eigenfunction distribute charge through currents in the electrolyte.

Multiplying by $1 - U_0^{(i)}(\eta)$ for $i \in \mathbb{N}^+$ and utilizing the relationship (89) yields

$$C^{(i)} = \frac{\int_0^1 \sum_{n=0}^{+\infty} B_n^{SS} P_{2n}(\eta) \left(1 - U_0^{(i)}(\eta)\right) \eta \mathrm{d}\eta}{\int_0^1 U_0^{(j)}(\eta) \left(1 - U_0^{(i)}(\eta)\right) \eta \mathrm{d}\eta}$$

$$= \frac{\frac{1}{\Lambda^{(i)}} \sum_{n=1}^{+\infty} \frac{\pi M'_{2n}(0)}{4(4n+1)} B_n^{(i)} B_n^{SS}}{\frac{\delta_{ij}}{\Lambda^{(i)}} \sum_{n=1}^{+\infty} \frac{\pi M'_{2n}(0)}{4(4n+1)} \left(B_n^{(i)}\right)^2} \quad (91)$$

$$= \frac{\left(\boldsymbol{b}_1^{SS}\right)^T \boldsymbol{M}_1 \boldsymbol{b}_1^{(i)}}{\left(\boldsymbol{b}_1^{(i)}\right)^T \boldsymbol{M}_1 \boldsymbol{b}_1^{(i)}} \quad .$$

Although the eigenfunctions of the transient components are independent of the Faradaic reaction, the decay time constants $\tau^{(i)}$ and the corresponding coefficients $C^{(i)}$ are not, as $\boldsymbol{b}_1^{SS}$ is dependent on $G$. With all the coefficients $C^{(i)}$ given for $i > 0$ (Table B6), the spatial distribution of the transient response is solved and the potential and current density distributions on the electrode surface at $t = 0^+$ are shown in Figure 15. The transient response constructed from eigenfunctions shows ripples due to the Gibbs effect (see Numeric Accuracy section), and therefore, the current distribution was spatially filtered (equiripple filter with the following parameters: passband frequency of 20 cycles per unit length, stopband frequency of 100 cycles per unit length, passband ripple of 1 dB, and stopband attenuation of 80 dB). The transient response is dominated by the lower order eigenfunctions. As the Faradaic conductance increases, the coefficients of $C^{(i)}$, $i \geq 1$ decreases. This indicates that more transient current runs through local Faradaic charge transfer versus through the electrolyte.

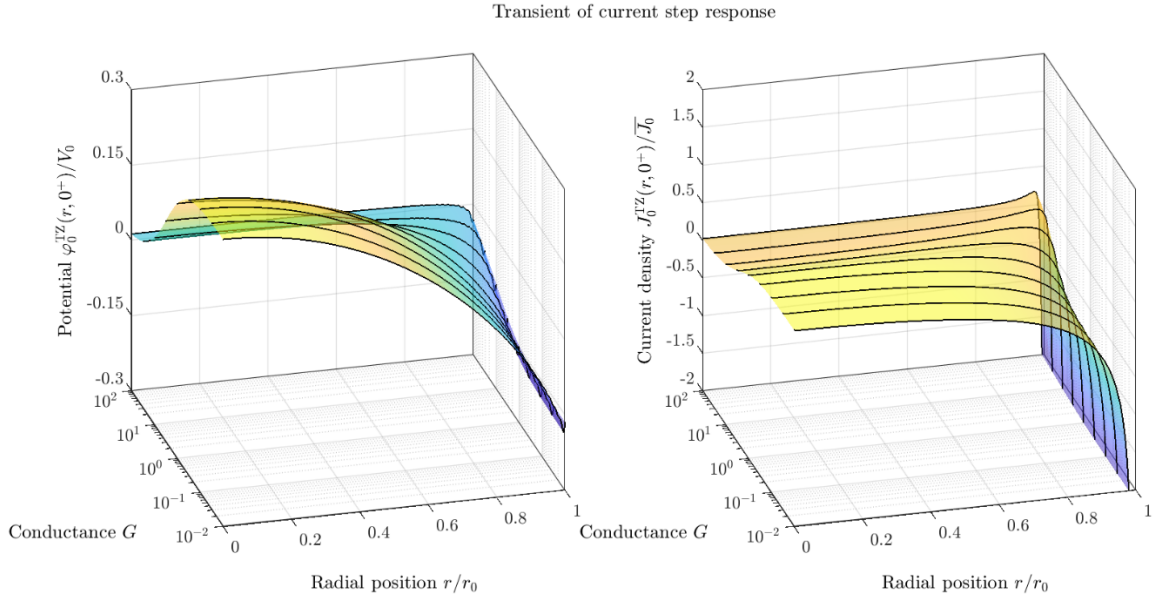

Figure 15. The potential and current density distribution of the transient response to current step input on the surface of the electrode at $t = 0^+$ for different Faradaic reaction conductance. The current density is filtered with a low pass filter.

For the electrode voltage, however, $C^{(0)}$ remains to be solved, and could be obtained by evaluating (61) and (78) at $\theta = 0$

$$C^{(0)} = \begin{cases} \dfrac{V^{\text{SS}}}{V_0} - 1 - \sum_{j=1}^{+\infty} C^{(j)} & , \quad G \neq 0 \\ 0 & , \quad G = 0 \end{cases}. \tag{92}$$

However, an easier way is through analysis of the current density components on the electrode-electrolyte interface. The steady state solution only contains Faradaic current as the double layer is charged to its asymptotic value. On the other hand, the transient response contains both Faradaic current and capacitive current component that are not equal in magnitude at any location. The initial condition is the primary distribution in the electrolyte at $t = 0^+$, and as the overpotential is zero, there is only capacitive current. Thus, the initial capacitive current of the transient response equals the primary current distribution on the electrode

$$\gamma \dfrac{\partial \left( V^{\text{TZ}}(t) - \varphi^{\text{TZ}}(r,z,t) \right)}{\partial t} \bigg|_{z=0, t=0^+} = -J_0^{\text{P}}(r,0) \quad, \tag{93}$$

which is

$$-\dfrac{\gamma}{\tau} V_0 \sum_{i=0}^{+\infty} C^{(i)} \left( \Lambda^{(i)} + G \right) \left( 1 - U_0^{(i)}(\eta) \right) = -\dfrac{2 \kappa V_0}{\pi \, r_0 \eta} \quad. \tag{94}$$

Multiplying both side by $\eta$ and integrating over 0 to 1, the summation equals zero for $i \neq 0$ utilizing (89), therefore yielding

$$C^{(0)}G = 1 \quad \Leftrightarrow \quad C^{(0)} = G^{-1} \quad . \tag{95}$$

Multiplying both side of (94) by $U_0^{(j)}(\eta)\eta$ and integrating over 0 to 1 also yields a computationally less expensive method for coefficients with $i \geq 1$

$$\begin{aligned}
C^{(i)} &= \frac{\Lambda^{(i)} \sum_{n=1}^{+\infty} A_{0,n} B_n^{(i)}}{2(\Lambda^{(i)} + G) \sum_{n=1}^{+\infty} \frac{-\pi M'_{2n}(0)}{4(4n+1)} \left(B_n^{(i)}\right)^2} \\
&= \frac{\Lambda^{(i)}}{2(\Lambda^{(i)} + G)\left(\boldsymbol{b}_1^{(j)}\right)^T \boldsymbol{M}_1 \boldsymbol{b}_1^{(j)}} \quad , \quad i \geq 1 \quad ,
\end{aligned} \tag{96}$$

which shows explicitly their dependency on $G$.

**Voltage of an ideally polarizable electrode**

In the special case of an ideally polarizable electrode, with $G = 0$, the electrode voltage $V(t)$ is unbounded. Intuitively, the current could only pass through the system by constantly charging the double layer capacitance. While the current density and potential distributions in the electrolyte will reach an asymptote, the electrode voltage and overpotential will continue to grow indefinitely. The current on the interface will shift from the primary distribution to a uniform distribution, and the boundary condition (11) becomes

$$-\frac{\kappa}{r_0 \eta} \left.\frac{\partial \Phi^{SS}(\eta,\xi)}{\partial \xi}\right|_{\xi=0+} = \bar{J}_0 = \frac{I_0}{\pi r_0^2} \quad . \tag{97}$$

The coefficients $B_n^{SS}$ could be calculated as

$$B_n^{SS} = -\frac{4(4n+1)}{\pi M'_{2n}(0)} a_{0,n} \tag{98}$$

or

$$\begin{aligned} \boldsymbol{b}_1^{SS} &= \boldsymbol{M}_1^{-1} \boldsymbol{a}_1 \\ \boldsymbol{b}_0^{SS} &= \boldsymbol{M}_0^{-1} \boldsymbol{a}_0 \end{aligned} \quad , \tag{99}$$

which could also be directly obtained from (67) and (70) with $G \to 0$.

The zeroth order term of the transient response is no longer transient and becomes part of the "steady state" response. In combination with the $G^{-1}$ term in (73), the zeroth order term becomes

$$\lim_{G \to 0}\left(G^{-1} + C^{(0)}e^{-\theta G}\right) = \lim_{G \to 0} \frac{\left(1 - e^{-\theta G}\right)}{G} = \theta \quad , \tag{100}$$

with $C^{(0)} = G^{-1}$.

Therefore, the "steady-state" voltage on an ideally polarizable electrode grows linearly as the step input is maintained

$$\frac{V^{\text{SS}}(t)}{V_0} = \theta + \frac{\overline{\varphi_0^{\text{SS}}}}{V_0} \quad , \tag{101}$$

with the average potential in the electrolyte above the electrode given as[1]

$$\frac{\overline{\varphi_0^{\text{SS}}}}{V_0} = 2\boldsymbol{a}_0^{\text{T}} \boldsymbol{M}_0^{-1} \boldsymbol{a}_0 = \sum_{n=0}^{+\infty} \frac{(4n+1) P_{2n}^4(0)}{(2n-1)^2 (n+1)^2} = \frac{32}{3\pi^2} \quad . \tag{102}$$

---

[1] This number with $\pi^2$ was given in the original article by Nisancioğlu and Newman (1973a, Eq. [39]). I remember being able to deduce it in the last step of the series summation during the initial writing of this review in 2012, however, it eluded me in later years. I would appreciate any tips and comments on the calculation of this series. —B. Wang.

## Transient Response to Voltage Step Input ([Nisancioğlu and Newman, 1973b](#))

The transient response to a voltage step input

$$V(t) = V_0 \cdot u(t) \tag{103}$$

applied to the electrode is similarly decomposed to a steady state solution and a transient solution.

$$\varphi(r, z, t) = \varphi^{SS}(r, z) \cdot u(t) - \varphi^{TZ}(r, z, t) \quad . \tag{104}$$

**Steady State Response**

The steady state solution for the voltage step input has a different current compared to the initial primary response $I^{SS} \neq I_0$, and the transient response contributes a net current $I^{TZ}(t)$ that results in the difference between the primary current value and the steady state value. Nevertheless, the steady state solution of this system could be easily given by scaling the results of the previous calculation for the current step input. Since the voltage is forced to stay at its initial value $V(t) \equiv V_0 = V^{SS}$, all the other steady state values ($\varphi^{SS}(r, z)$, $J_0^{SS}(r)$, $I^{SS}$, etc.) will be scaled by a factor

$$K = \left(G^{-1} + 2\boldsymbol{a}_0^T \big(G(\boldsymbol{A}_0 - 2\boldsymbol{a}_0\boldsymbol{a}_0^T) + \boldsymbol{M}_0\big)^{-1} \boldsymbol{a}_0 \right)^{-1} , \tag{105}$$

which is the inverse of (73). The scaling factor is plotted in Figure 16 as a function of the Faradaic reaction $G$. If $G = 0$, then this scaling factor becomes zero, and all the steady state values in the electrolyte are also zero. This is intuitive as the steady state without Faradaic reaction to discharge the double layer capacitance results in fully charged capacitance with no current following in the electrolyte. The steady state potential and current density distributions on the electrode surface is shown in Figure 17.

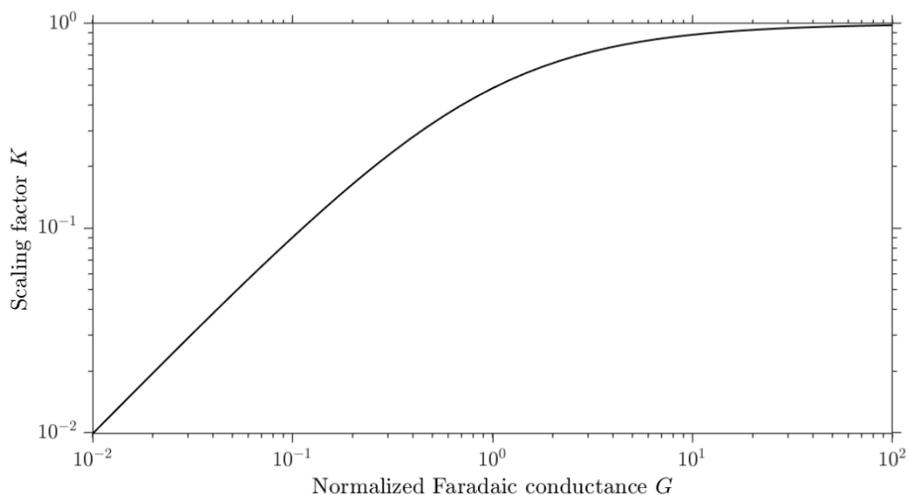

Figure 16. Scaling factor the steady state response of the voltage step input compared to that of the current step input as a function of Faradaic reaction conductance.

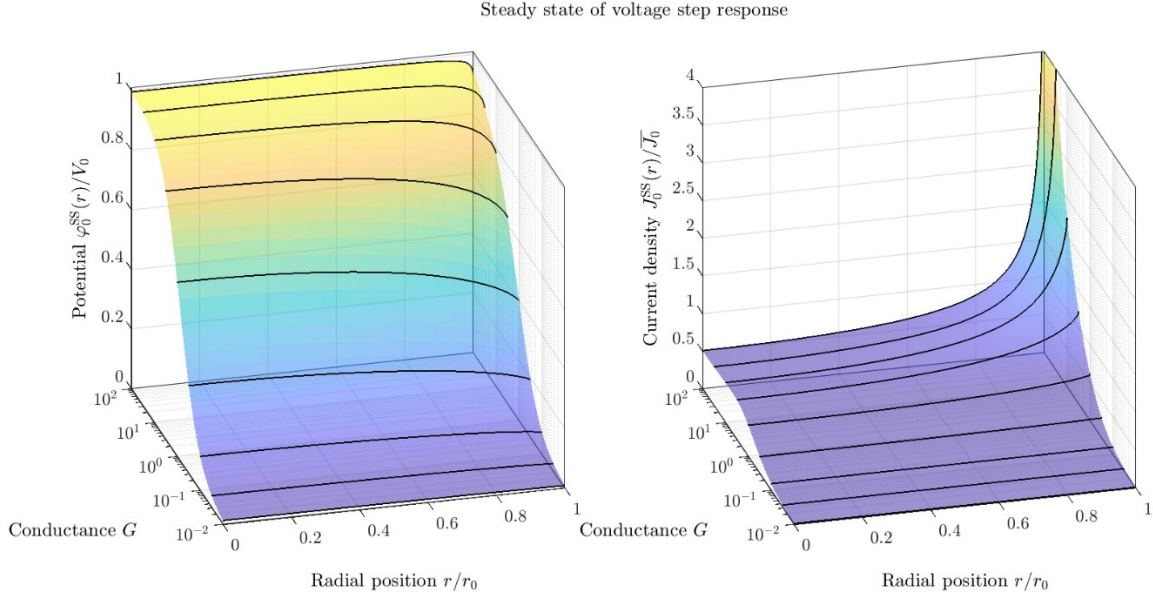

Figure 17. The steady state potential and current density distribution on the surface of the electrode for different Faradaic reaction conductance.

**Eigenfunctions of the Transient Response**

In the voltage step input situation, the transient response will follow a different course to connect the primary distribution to the steady state and is the focus of the analysis. Using the same notation in (75), the transience is given as

$$\frac{\varphi^{\mathrm{TZ}}(\eta,\xi,\theta)}{V_0} = \left[\sum_{i=0}^{+\infty} C^{(i)} e^{-\theta(\Lambda^{(i)}+G)} U^{(i)}(\eta,\xi)\right] \cdot u(\theta) \quad, \tag{106}$$

with the summation starting from 0, indicating the net current. The boundary condition (11) is

$$\left.\frac{\partial U^{(i)}(\eta,\xi)}{\partial \xi}\right|_{\xi=0+} + \frac{4\eta}{\pi}\Lambda^{(i)} U_0^{(i)}(\eta) = 0 \quad. \tag{107}$$

Using the same definition for $U^{(i)}$ as in (77) yields

$$\sum_{n=0}^{+\infty}\left[\Lambda^{(i)} a_{m,n} + \frac{\pi M'_{2m}(0)\delta_{mn}}{4(4m+1)}\right] B_n^{(i)} = 0 \quad, \quad m \in \mathbb{N}^0 \quad, \tag{108}$$

which, when rewritten in matrix form

$$\left(\boldsymbol{M}_0 - \Lambda^{(i)} \boldsymbol{A}_0\right) \boldsymbol{b}_0^{(i)} = 0 \quad, \tag{109}$$

shows that $\Lambda^{(i)}$ and $\boldsymbol{b}_0^{(i)}$ are eigenvalues and eigenvectors of matrix $\boldsymbol{M}_0(\boldsymbol{A}_0)^{-1}$, respectively. All eigenvalues, except for $\Lambda^{(0)}$, have a counterpart similar in value for the current step and voltage step input problem. The zeroths coefficients of $\boldsymbol{b}_0^{(i)}$ are set to $B_0^{(i)} = 1$ for normalization, and the eigenfunctions are scaled by the coefficients $C^{(i)}$. Applying the same technique as in (82)–(85) yields the equation for the remaining coefficients

$$\boldsymbol{b}_1^{(i)} = -\left(\boldsymbol{A}_1 - \frac{\boldsymbol{M}_1}{\Lambda^{(i)}}\right)^{-1} \boldsymbol{a}_1 \quad , \tag{110}$$

which is identical to (88) for the current step response except for the sign. See Table B7 for the numeric values of $\Lambda^{(i)}$ and $\boldsymbol{b}_0^{(i)}$ obtained with $n_{\max} = 1024$.

The potential and current density distributions of the eigenfunctions are shown in Figure 18. The current density on the electrode is proportional to $\varphi_0^{(i)}(\eta, \theta)$ or $U_0^{(i)}(\eta)$ according to (107). The eigenfunctions therefore satisfy an orthogonality described as

$$\int_0^1 U_0^{(i)}(\eta) U_0^{(j)}(\eta) \eta \mathrm{d}\eta = \frac{\delta_{ij}}{\Lambda^{(i)}} \sum_{n=0}^{+\infty} \frac{-\pi M_{2n}'(0)}{4(4n+1)} \left(B_n^{(i)}\right)^2 \\ = \frac{\delta_{ij}}{\Lambda^{(i)}} \left(\boldsymbol{b}_0^{(i)}\right)^T \boldsymbol{M}_0 \boldsymbol{b}_0^{(i)} \quad . \tag{111}$$

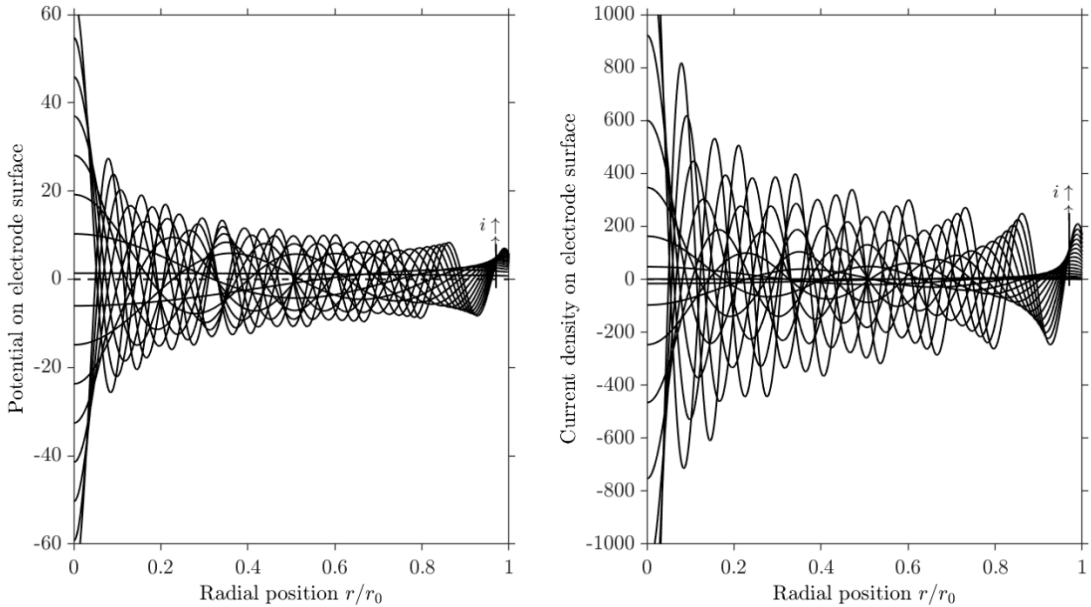

Figure 18. The first 16 normalized and dimensionless "potential" and "current density" distribution on the surface of the electrode of the eigenfunctions of the transient response to voltage step input.

**Transient Response**

Similar to (91), the coefficients $C^{(i)}$ could be determined by the initial condition of the potential. The coefficients are given by

$$C^{(i)} = \frac{\left(\boldsymbol{b}_0^{\text{SS}}\right)^T \boldsymbol{M}_0 \boldsymbol{b}_0^{(i)} - 1/2}{\left(\boldsymbol{b}_0^{(i)}\right)^T \boldsymbol{M}_0 \boldsymbol{b}_0^{(i)}} \quad , \tag{112}$$

in which $\boldsymbol{b}_0^{\text{SS}}$ is scaled by $K$ compared to its counterpart of the current step response. Or utilizing the same analysis of the current density components, the same conclusion holds for the potential step input, i.e. the initial capacitive current of the transient solution equals the primary current distribution on the electrode:

$$\gamma \frac{\partial \left(0 - \varphi^{\text{TZ}}(r,z,t)\right)}{\partial t}\bigg|_{z=0, t=0^+} = -J_0^{\text{P}}(r,0) \quad , \tag{113}$$

which is

$$-\frac{\gamma}{\tau} V_0 \sum_{i=0}^{+\infty} C^{(i)} \left(\Lambda^{(i)} + G\right) U_0^{(i)}(\eta) = -\frac{2}{\pi} \frac{\kappa V_0}{a\eta} \quad . \tag{114}$$

Using (111) yields an alternative and computationally less expensive expression

$$C^{(i)} = \frac{\Lambda^{(i)}}{2(\Lambda^{(i)} + G) \sum_{n=0}^{+\infty} \frac{\pi M'_{2n}(0)}{4(4n+1)} \left(B_n^{(i)}\right)^2} = \frac{-\Lambda^{(i)}}{2(\Lambda^{(i)} + G) \left(\boldsymbol{b}_0^{(i)}\right)^T \boldsymbol{M}_0 \boldsymbol{b}_0^{(i)}} \quad , \tag{115}$$

which is consistent with (96) for the current step response. Numeric values of $C^{(i)}$ are given in Table B8.

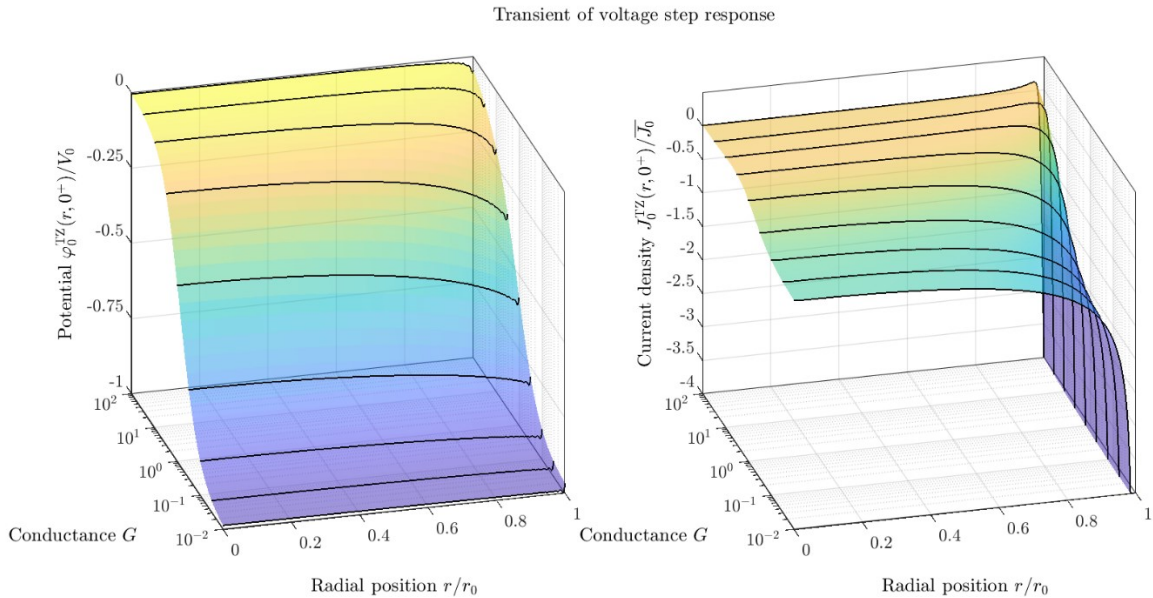

Figure 19. The potential and current density distribution of the transient response to voltage step input on the surface of the electrode at $t = 0^+$ for different Faradaic reaction conductance. The traces were filtered after the construction from the eigenfunction.

The transient response is thus solved, and the potential and current density distributions on the electrode surface at $t = 0^+$ are shown in Figure 19. With $B_0^{(i)} = 1$, the electrode current is given as

$$\frac{I(\theta)}{I_0} = \frac{I^{SS}}{I_0} - \sum_{n=0}^{+\infty} C^{(i)} e^{-\theta(\Lambda^{(i)}+G)} \quad . \tag{116}$$

**Numeric Accuracy**

Numeric solution for the frequency dispersion, current step response and voltage step response all involve an infinite set of equations, summarized in the form of matrices. The matrices need to be truncated for the solution to be possible, and therefore the accuracy of the solutions should be evaluated.

To solve the coefficients $\boldsymbol{b}_0^H$ for the frequency dispersion, (53) involves inverse operation of a matrix. However, as the terms of $\boldsymbol{a}_0$ on the right hand side of the equation decrease with increasing $n$, the accuracy of the calculated terms $\boldsymbol{b}_0^H$ could be guaranteed by setting $n_{max}$ so that the $n$th term of $\boldsymbol{a}_0$, i.e. $a_{0,n}$ is small enough. For the calculation, $n_{max}$ was set to 1024, which has an accuracy of $10^{-12}$ (relative difference when changing $n_{max}$ by 1). Also, the coefficients $B_n^H$ decay fast in amplitude, allowing accurate calculation of the potential and current density distribution with only a few basis functions.

For the current step response and voltage step response, the truncation of (87) and (109) reduces the accuracy of the eigenvalues. To investigate which of the eigenvalues are accurate, the eigenvalues are plotted from solutions obtained with several $n_{max}$ (8, 16, 32, 64, 128, 256, 512, and 1024), showing high linearity (Figure 20). The separation of the eigenvalues appears to be $\pi^2/4$.

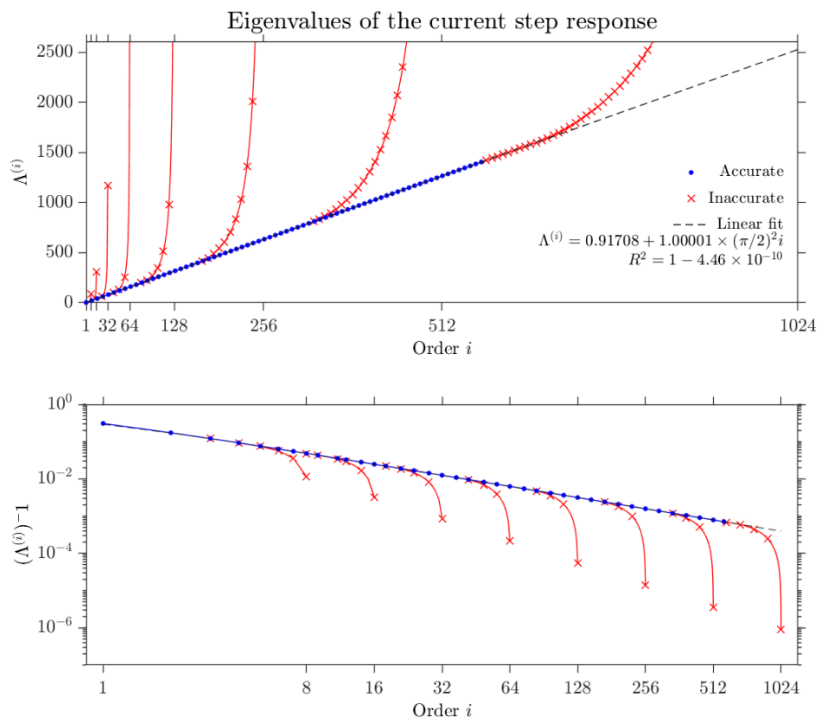

Figure 20. The eigenvalues of the current step input solved with different size of matrix truncation. Markers in blue show eigenvalues considered accurate after comparing the values solved from different sizes of matrix truncation. Markers in red show eigenvalues considered inaccurate when comparing with those solved by matrix truncated to larger size. The dashed line is a linear regression of the accurate eigenvalues, showing highly linear behavior.

The accuracy could be checked by comparing the same order eigenvalues calculated with different $n_{\max}$, with a relative difference less than 0.01% being considered accurate. Apparently, as $n_{\max}$ increases, more eigenvalues become accurate enough to be included for later calculations and the corresponding coefficients belong to the "believable scales" (Boyd, 2001, p. 427). Approximately 50%–60% of the eigenvalues and coefficients can be considered accurate, which is an empirical rule for "spectral blocking" (Boyd, 2001, pp. 132, 207, and 427). Similar results could be obtained for the voltage step response, as shown in Figure 21, with the same separation of $\pi^2/4$.[2]

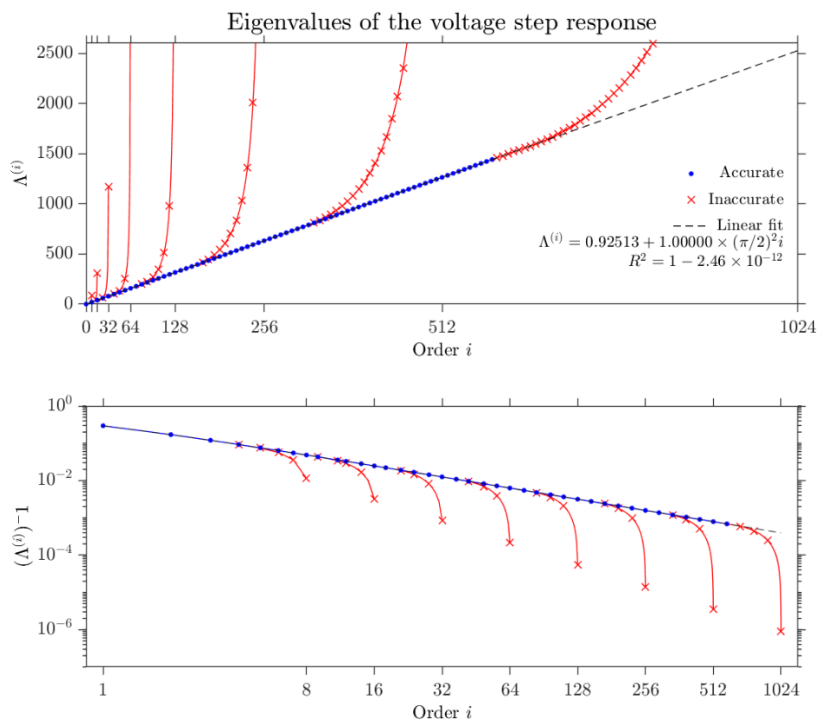

Figure 21. The eigenvalues of the voltage step input, plotted in the same format as Figure 20, expect with order starting at 0 instead of 1.

Another aspect of accuracy arises when assembling the current density distribution on the electrode surface from the basis functions due to its behavior at the disk's edge. Figure 22 shows Gibbs ripple towards the edge of the electrode, as often seen in the reconstruction of Fourier series. For small $n_{\max}$, the ripples could be somewhat reduced with increased $n_{\max}$ (Antohi and Scherson, 2006), but these "spectral ringing" are inherent due to the discontinuity and independent of $n_{\max}$ and therefore should be carefully treated

---

2　The high linearity and same equal spacing involving $\pi^2$ are unlikely coincidences. I suspect that they can be proven mathematically but are beyond my capabilities. —B. Wang ([Boyd, 2001](), p. 419). The current density distribution presented in the previous sections were spatially filtered to get rid of the oscillation with frequencies higher than 20 cycles per unit length, while keeping in mind that ideally there is a singularity at the very edge. An alternative way to obtain the current density without such ripples is to utilize the reconstructed potential distribution, which is continuous and should not suffer from Gibbs phenomenon.

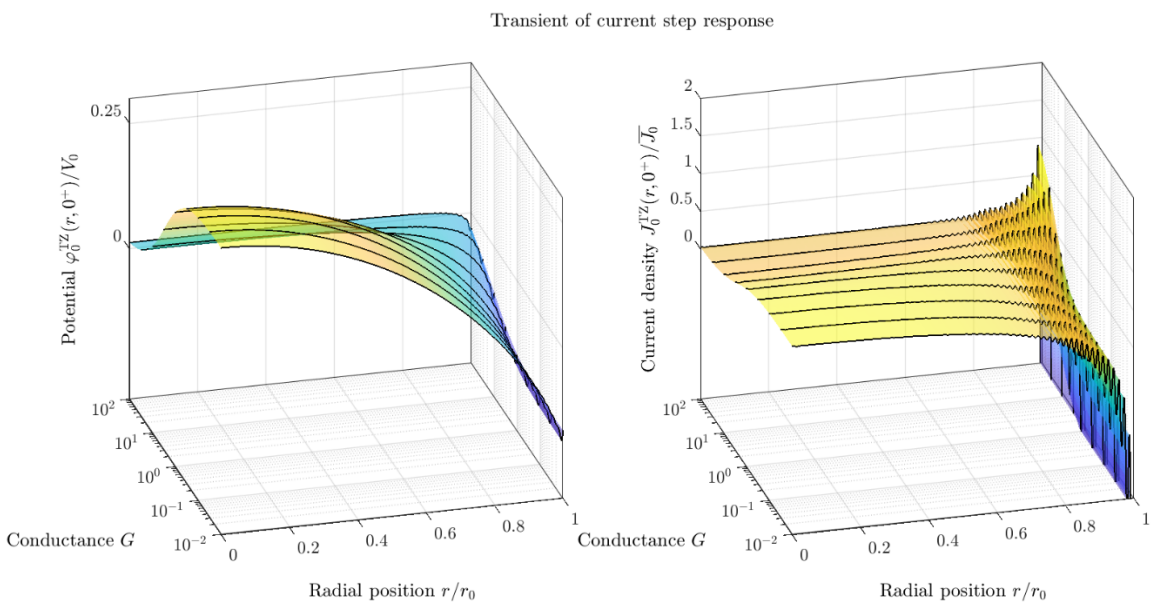

Figure 22. The unprocessed current density distribution of the transient response to current step input on the surface of the electrode at $t = 0^+$ for a range charge transfer conductance, corresponding to Figure 15. The Gibbs ripples demonstrate the spectral ringing at the electrode's edge due to the current density reaching infinity.

# Appendix A—Legendre Functions and their Extension on the Imaginary Axis

To solve (22), the Legendre functions need to be extended to the imaginary axis. The Legendre function are first introduced, with background knowledge only related to this review given. For further details on the Legendre function, any mathematical textbook can be consulted. The detailed derivation of the solution to (22) that Newman (1966b) omitted is then discussed to complete the analysis on the Basis Functions.

**Legendre Functions**

Legendre functions are solution to the Legendre equation, a second order differential equation derived from Laplace's equation in a spherical coordinate system $(r, \phi, \theta)$. Using separation of variables, the equation for $\theta = \arccos x$ under axial symmetric condition is the Legendre equation

$$\frac{d}{dx}\left[(1-x^2)\frac{df(x)}{dx}\right] + l(l+1)f(x) = 0 \quad , \quad |x| \leq 1 \quad . \tag{A1}$$

The solution to the Legendre equation is

$$f(x) = c_1 P_l(x) + c_2 Q_l(x) \quad , \tag{A2}$$

with $P_l(x)$ and $Q_l(x)$ being the Legendre function of the first and second kind, respectively. Generally speaking, $l$ could be complex, and usually $P_l(x)$ and $Q_l(x)$ are complex and not bounded at the points $x = \pm 1$. If $l \in \mathbb{N}^0$, then $P_l(x)$ becomes a polynomial and with normalization $P_l(1) = 1$:

$$P_l(x) = \sum_{m=0}^{\lfloor l/2 \rfloor} \frac{(-1)^m (2l-2m)!}{2^l \cdot m! \, (l-m)! \, (l-2m)!} x^{l-2m} \quad . \tag{A3}$$

$Q_l(x)$ is obtained from $P_l(x)$ by the method of reduction of order:

$$Q_l(x) = u_l(x) P_l(x) \quad , \tag{A4}$$

with

$$u_l(x) = \int^x \frac{d\tilde{x}}{(1-\tilde{x}^2) P_l^2(\tilde{x})} \quad . \tag{A5}$$

$Q_l(x)$ is unbounded at $x = \pm 1$ and could also be written as

$$Q_l(x) = P_l(x) Q_0(x) - W_l(x) \quad , \tag{A6}$$

where $W_l(x)$ denotes a polynomial of order $l-1$.

Legendre functions have several characteristics, with the following two being important for the disk electrode analysis:

- Orthogonality: Legendre functions are orthogonal polynomials over the interval $[-1,1]$:

$$\int_{-1}^{1} P_n(x) P_m(x) \, dx = \frac{2}{2l+1} \delta_{nm} \quad . \tag{A7}$$

- Recurrent generation: Higher order Legendre functions could be generated by lower order ones (also applies to $Q_l(x)$ and $W_l(x)$):

$$(l+1)P_{l+1}(x) = (2l+1)xP_l(x) - lP_{l-1}(x) \quad . \tag{A8}$$

Starting functions for recurrent generation are given as follows:

$$\begin{cases} P_0(x) = 1 \\ P_1(x) = x \\ P_2(x) = \dfrac{3x^2 - 1}{2} \end{cases}, \quad \begin{cases} Q_0(x) = \dfrac{1}{2}\ln\left(\dfrac{1+x}{1-x}\right) \\ Q_1(x) = \dfrac{x}{2}\ln\left(\dfrac{1+x}{1-x}\right) - 1 \\ Q_2(x) = \dfrac{(3x^2 - 1)}{4}\ln\left(\dfrac{1+x}{1-x}\right) - \dfrac{3x}{2} \end{cases}, \quad \begin{cases} W_0(x) = 0 \\ W_1(x) = 1 \\ W_2(x) = \dfrac{3x}{2} \end{cases} \quad . \tag{A9}$$

The lower order functions are given in Figure A1 for even and odd number of $l$.

**Evaluating Legendre Functions on the Imaginary Axis**

Substituting the variable $\xi$ by a purely imaginary one

$$\xi = f(\hat{\xi}) = i\hat{\xi} \quad , \tag{A10}$$

the function $M(\xi)$ satisfies

$$\begin{cases} M(\xi) = M(f(\hat{\xi})) = \widehat{M}(\hat{\xi}) \\ \dfrac{d\widehat{M}(\hat{\xi})}{d\hat{\xi}} = \dfrac{dM(\xi)}{d\xi}\dfrac{d\xi}{d\hat{\xi}} = i\dfrac{dM(\xi)}{d\xi} \end{cases} \quad . \tag{A11}$$

Therefore for each $\lambda = l(l+1)$, substituting (A11) into (22) gives

$$\frac{d}{d\hat{\xi}}\left[(1-\hat{\xi}^2)\frac{d\widehat{M}(\hat{\xi})}{d\hat{\xi}}\right] + l(l+1)\widehat{M}(\hat{\xi}) = 0 \quad . \tag{A12}$$

Equation (A12) indicates that $\widehat{M}(\hat{\xi})$ is a Legendre function, hence

$$\widehat{M}_l(\hat{\xi}) = c_l^{MP}P_l(\hat{\xi}) + c_l^{MQ}Q_l(\hat{\xi}) \quad , \tag{A13}$$

and

$$M_l(\xi) = \widehat{M}_l(\xi/i) = c_l^{MP}P_l(\xi/i) + c_l^{MQ}Q_l(\xi/i) \quad . \tag{A14}$$

Before the Legendre functions are evaluated on the imaginary axis, the variable is first extended to the complex plane $z = x + i \cdot y \in \mathbb{C}$, and $P_l(z)$ and $Q_l(z)$ become complex. With $l \in \mathbb{N}^0$, $P_l(z)$ is a polynomial, and $Q_l(z)$ could also be obtained by (A4)–(A6). The corresponding regions of convergence (ROC) are $z \in \mathbb{C}\setminus\{z = \infty\}$ for $P_l(z)$, and $z \in \mathbb{C}\setminus\{z = \pm 1, \infty\}$ for $Q_l(z)$, respectively. The latter could be inferred from the recurrent generation of $Q_l(z)$ and the expression of the first term $Q_0(z)$

$$Q_0(z) = \frac{1}{4}\ln\left[\frac{(1+x)^2 + y^2}{(1-x)^2 + y^2}\right] + \frac{i}{2}\operatorname{atan}\left(\frac{2y}{1-x^2-y^2}\right) \quad . \tag{A15}$$

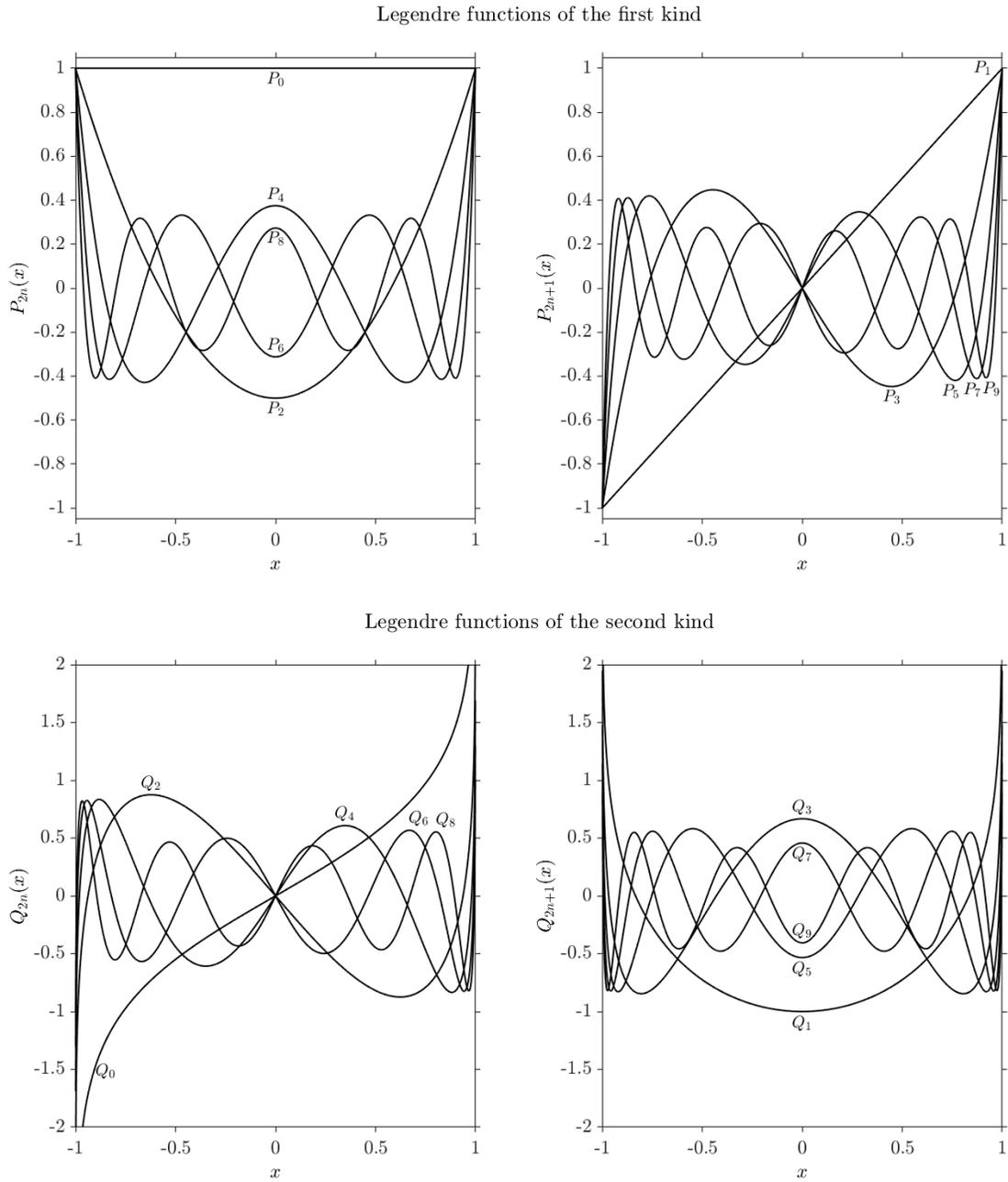

Figure A1. Legendre functions of the first and second kind, plotted separately for even and odd orders. For $P_l(x)$, the oscillations have amplitude of $O(n^{-1/2})$ over most of the interval, with the a narrow boundary layers near the endpoints where the polynomial rises to $\pm 1$ (Boyd, 2001, p53).

Modifying (A4)–(A6) gives

$$\frac{Q_l(\xi/\mathrm{i})}{P_l(\xi/\mathrm{i})} = u_l(\xi/\mathrm{i}) = \int^{\xi/\mathrm{i}} \frac{\mathrm{d}\tilde{x}}{(1-\tilde{x}^2)P_l^2(\tilde{x})} = -\mathrm{i}\int^{\xi} \frac{\mathrm{d}\hat{x}}{(1+\hat{x}^2)P_l^2(\hat{x}/\mathrm{i})} \quad , \tag{A16}$$

$$Q_l(\xi/\mathrm{i}) = P_l(\xi/\mathrm{i})Q_0(\xi/\mathrm{i}) - W_l(\xi/\mathrm{i}) \quad . \tag{A17}$$

The functions $P_l(\xi/i)$ and $Q_l(\xi/i)$ are bounded on $\xi \in [0, +\infty)$, and unbounded at $+\infty$. $l = 0$ is the only exception where both functions converge for $\xi \to +\infty$.

The first few functions are given as

$$\begin{cases} P_0(\xi/i) = 1 \\ P_1(\xi/i) = \xi/i \\ P_2(\xi/i) = -\dfrac{3\xi^2 + 1}{2} \end{cases}, \begin{cases} Q_0(\xi/i) = -i\,\mathrm{atan}(\xi) \\ Q_1(\xi/i) = -(\xi\,\mathrm{atan}(\xi) + 1) \\ Q_2(\xi/i) = \dfrac{i}{2}[(3\xi^2 + 1)\,\mathrm{atan}(\xi) + 3\xi] \end{cases}, \begin{cases} W_0(\xi/i) = 0 \\ W_1(\xi/i) = 1 \\ W_2(\xi/i) = -\dfrac{3i\xi}{2} \end{cases}. \quad (A18)$$

With $l = 2n$, $P_{2n}(\xi/i)$ is real and $Q_{2n}(\xi/i)$ and $W_{2n}(\xi/i)$ are imaginary. For $M_{2n}(\xi)$ to be real, $c_{2n}^{MQ}$ is imaginary. To satisfy the conditions relevant to $M_{2n}(\xi)$

$$\begin{cases} |M_{2n}(\xi)| < +\infty \\ \lim_{\xi \to +\infty} M_{2n}(\xi) = 0 \\ M_{2n}(0) = 1 \end{cases}. \quad (A19)$$

the following must be true:

$$\begin{cases} M_{2n}(0) = c_{2n}^{MP} P_{2n}(0) = 1 \\ \lim_{\xi \to +\infty} c_{2n}^{MP} P_{2n}(\xi/i) + c_{2n}^{MQ} Q_{2n}(\xi/i) = 0 \end{cases} \Rightarrow \begin{cases} c_{2n}^{MP} = \dfrac{1}{P_{2n}(0)} \\ c_{2n}^{MQ} = -c_{2n}^{MP} \lim_{\xi \to +\infty} \left( \dfrac{P_{2n}(\xi/i)}{Q_{2n}(\xi/i)} \right) \end{cases}. \quad (A20)$$

The coefficient $c_{2n}^{MP}$ (Table B1) could be easily calculated as the reciprocal of $P_{2n}(0)$:

$$c_{2n}^{MP} = \frac{1}{P_{2n}(0)} = \frac{(-1)^n (2^n \cdot n!)^2}{(2n)!}, \quad (A21)$$

which can be shown to asymptotically scale with the square root of $n$

$$\lim_{n \to +\infty} c_{2n}^{MP} = (-1)^n \sqrt{\pi n}, \quad (A22)$$

by using the Stirling's approximation of factorials for large $n$

$$n! \approx \sqrt{2\pi n} \left(\frac{n}{e}\right)^n. \quad (A23)$$

The limit for the coefficient $c_{2n}^{MQ}$ could be calculated from (A16) as

$$\lim_{\xi \to +\infty} \left( \frac{P_{2n}(\xi/i)}{Q_{2n}(\xi/i)} \right) = \left[ \lim_{\xi \to +\infty} \left( Q_0(\xi/i) + \frac{W_{2n}(\xi/i)}{P_{2n}(\xi/i)} \right) \right]^{-1} = \frac{2i}{\pi}, \quad (A24)$$

and therefore

$$c_{2n}^{MQ} = -\frac{2i}{\pi} c_{2n}^{MP} = -\frac{2i}{\pi} \frac{(-1)^n (2^n \cdot n!)^2}{(2n)!}. \quad (A25)$$

The functions $M_{2n}(\xi)$ are shown on logarithmic scale in Figure A2. It should be noticed that the decay to zero is very fast even for small $n$. For example, for $n \geq 1$ the function decays to less than $10^{-2}$ for $\xi \leq$

3. Hence the zeroth order of the solution will dominate the middle to far field in the electrolyte, and the electrode would be perceived as a point source from far away. Higher order functions are decaying extremely fast and any contribution would be very limited to the origin of the $\xi$ axis, i.e. very close to the electrode-electrolyte interface.

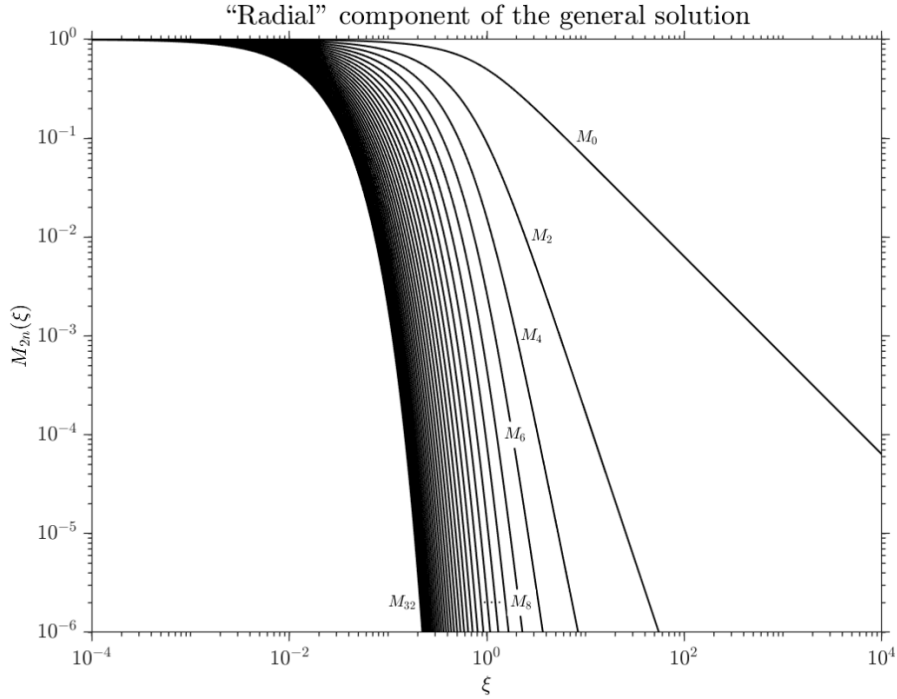

Figure A2. The "radial" component of the solution $M_{2n}(\xi)$ on log-log scale, obtained from evaluating the Legendre function on the imaginary axis.

The derivative of $M_{2n}(\xi)$ at the origin is used several times such as for calculating the current density on the electrode surface (27) and the matrix $\boldsymbol{M}_0$ (50). Utilizing the relationship (A16) and $P'_{2n}(0) = 0$, $M'_{2n}(0)$ is represented by the first term in the series expansion of $Q_{2n}(\xi/i)$

$$M'_{2n}(0) = c^{MQ}_{2n} \left.\frac{dQ_{2n}(\xi/i)}{d\xi}\right|_{\xi=0} = \frac{-2}{\pi P^2_{2n}(0)} = -\frac{2}{\pi} \frac{(2^n \cdot n!)^4}{[(2n)!]^2} \quad , \tag{A26}$$

and scales linearly with large $n$

$$\lim_{n \to +\infty} M'_{2n}(0) = \lim_{n \to +\infty} -\frac{2}{\pi} \frac{(2^n \cdot n!)^4}{[(2n)!]^2} = -2n \quad . \tag{A27}$$

Therefore, the diagonal elements of matrices $\boldsymbol{M}_0$ and $\boldsymbol{M}_1$ approach a constant

$$\lim_{n \to +\infty} m_{n,n} = \lim_{n \to +\infty} \frac{-\pi M'_{2n}(0)}{4(4n+1)} = \frac{\pi}{8} \quad . \tag{A28}$$

# Appendix B—Tables and Figures of Numerical Calculations

The data, figures, and MATLAB code for generating the data are available at: 10.5281/zenodo.4291332

**Legendre functions related results**

Table B1: Coefficient for constructing the Legendre functions on the imaginary axis. The first 50 of $c_{2n}^{\mathrm{MP}}$ are given, where as $c_{2n}^{\mathrm{MQ}}$ and $M'_{2n}(0)$ could be easily calculated from $c_{2n}^{\mathrm{MP}}$ by using (A25) and (A26). Full data for $n \leq 1024$ available online.

| $n$ | $c_{2n}^{\mathrm{MP}}$ | $n$ | $c_{2n}^{\mathrm{MP}}$ | $n$ | $c_{2n}^{\mathrm{MP}}$ |
|---|---|---|---|---|---|
| 0 | 1.00000 | 17 | −7.36194 | 34 | 10.37316 |
| 1 | −2.00000 | 18 | 7.57228 | 35 | −10.52349 |
| 2 | 2.66667 | 19 | −7.77694 | 36 | 10.67171 |
| 3 | −3.20000 | 20 | 7.97635 | 37 | −10.81790 |
| 4 | 3.65714 | 21 | −8.17089 | 38 | 10.96214 |
| 5 | −4.06349 | 22 | 8.36091 | 39 | −11.10450 |
| 6 | 4.43290 | 23 | −8.54671 | 40 | 11.24507 |
| 7 | −4.77389 | 24 | 8.72855 | 41 | −11.38390 |
| 8 | 5.09215 | 25 | −8.90669 | 42 | 11.52105 |
| 9 | −5.39169 | 26 | 9.08133 | 43 | −11.65659 |
| 10 | 5.67546 | 27 | −9.25268 | 44 | 11.79058 |
| 11 | −5.94572 | 28 | 9.42091 | 45 | −11.92305 |
| 12 | 6.20423 | 29 | −9.58618 | 46 | 12.05408 |
| 13 | −6.45240 | 30 | 9.74866 | 47 | −12.18369 |
| 14 | 6.69138 | 31 | −9.90848 | 48 | 12.31194 |
| 15 | −6.92212 | 32 | 10.06575 | 49 | −12.43887 |
| 16 | 7.14541 | 33 | −10.22061 | ⋮ | ⋮ |

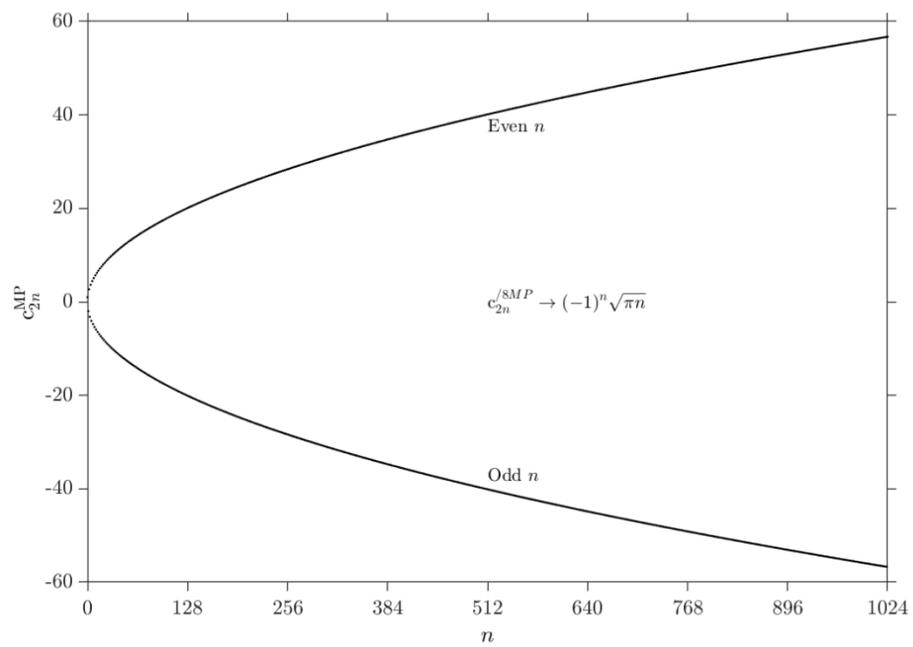

Figure B1. $c_{2n}^{\text{MP}}$ for $n \leq 1024$.

**Numeric matrices**

Table B2: $\boldsymbol{A_0}$ and $a_{0,0}$, $\boldsymbol{a_0}$, $\boldsymbol{a_1}$ and $\boldsymbol{A_1}$, in the form of $\boldsymbol{A_0} = \begin{bmatrix} a_0 & \boldsymbol{a_1^T} \\ & \boldsymbol{A_1} \end{bmatrix} = \begin{bmatrix} a_{0,0} & \boldsymbol{a_1^T} \\ \boldsymbol{a_1} & \boldsymbol{A_1} \end{bmatrix}$. Full data for $n, m \leq 1024$ available online.

|   | 0 | 1 | 2 | 3 | 4 | 5 | 6 | 7 | 8 | 9 | ... |
|---|---|---|---|---|---|---|---|---|---|---|---|
| 0 | 0.5000 | 0.1250 | −0.0208 | 0.0078 | −0.0039 | 0.0023 | −0.0015 | 0.0010 | −0.0007 | 0.0005 | ... |
| 1 | 0.1250 | 0.1250 | 0.0339 | −0.0063 | 0.0025 | −0.0013 | 0.0008 | −0.0005 | 0.0004 | −0.0003 | ... |
| 2 | −0.0208 | 0.0339 | 0.0703 | 0.0202 | −0.0037 | 0.0015 | −0.0008 | 0.0005 | −0.0003 | 0.0002 | ... |
| 3 | 0.0078 | −0.0063 | 0.0202 | 0.0488 | 0.0145 | −0.0027 | 0.0011 | −0.0006 | 0.0004 | −0.0002 | ... |
| 4 | −0.0039 | 0.0025 | −0.0037 | 0.0145 | 0.0374 | 0.0113 | −0.0021 | 0.0009 | −0.0005 | 0.0003 | ... |
| 5 | 0.0023 | −0.0013 | 0.0015 | −0.0027 | 0.0113 | 0.0303 | 0.0093 | −0.0018 | 0.0007 | −0.0004 | ... |
| 6 | −0.0015 | 0.0008 | −0.0008 | 0.0011 | −0.0021 | 0.0093 | 0.0254 | 0.0079 | −0.0015 | 0.0006 | ... |
| 7 | 0.0010 | −0.0005 | 0.0005 | −0.0006 | 0.0009 | −0.0018 | 0.0079 | 0.0219 | 0.0069 | −0.0013 | ... |
| 8 | −0.0007 | 0.0004 | −0.0003 | 0.0004 | −0.0005 | 0.0007 | −0.0015 | 0.0069 | 0.0193 | 0.0061 | ... |
| 9 | 0.0005 | −0.0003 | 0.0002 | −0.0002 | 0.0003 | −0.0004 | 0.0006 | −0.0013 | 0.0061 | 0.0172 | ... |
| ⋮ | ⋮ | ⋮ | ⋮ | ⋮ | ⋮ | ⋮ | ⋮ | ⋮ | ⋮ | ⋮ | ⋱ |

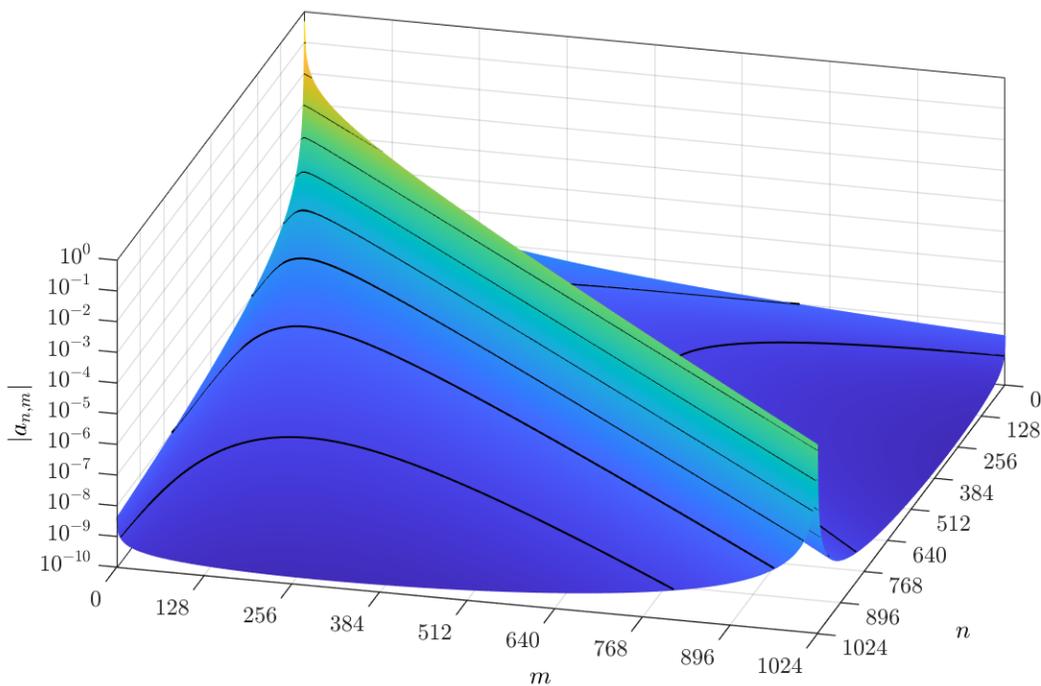

Figure B2. Surface plot of the absolute values of $\boldsymbol{A_0}$ for $n, m \leq 1024$. The off-diagonal entries are much smaller than the diagonal.

Table B3: $M_0$ and $M_1$ are diagonal matrices, with $M_0 = \begin{bmatrix} 1/2 & \mathbf{0}_1^T \\ \mathbf{0}_1 & M_1 \end{bmatrix}$. Full data for $n \leq 1024$ available online.

| $n$ | $m_{n,n}$ | $n$ | $m_{n,n}$ | $n$ | $m_{n,n}$ |
|---|---|---|---|---|---|
| 0 | 0.50000 | 11 | 0.39280 | 22 | 0.39272 |
| 1 | 0.40000 | 12 | 0.39278 | 23 | 0.39272 |
| 2 | 0.39506 | 13 | 0.39277 | 24 | 0.39272 |
| 3 | 0.39385 | 14 | 0.39276 | 25 | 0.39272 |
| 4 | 0.39337 | 15 | 0.39275 | 26 | 0.39272 |
| 5 | 0.39314 | 16 | 0.39275 | 27 | 0.39272 |
| 6 | 0.39301 | 17 | 0.39274 | 28 | 0.39271 |
| 7 | 0.39293 | 18 | 0.39274 | 29 | 0.39271 |
| 8 | 0.39288 | 19 | 0.39273 | 30 | 0.39271 |
| 9 | 0.39284 | 20 | 0.39273 | ⋮ | ⋮ |
| 10 | 0.39282 | 21 | 0.39273 | $+\infty$ | $\pi/8$ |

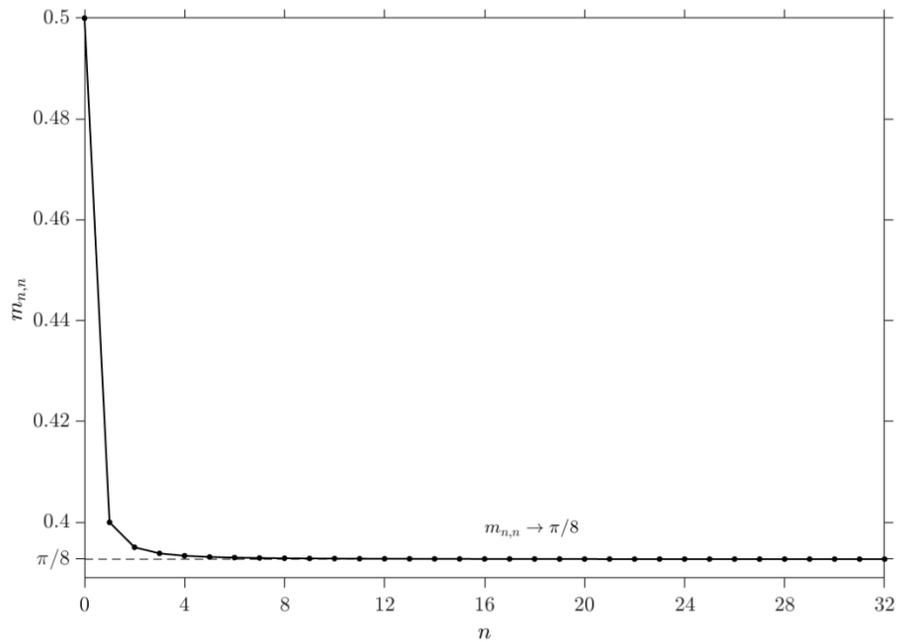

Figure B3. $m_{n,n}$ for $n \leq 32$. Convergences of $m_{n,n}$ towards its limit is very fast.

**Current Step Response**

Table B4: Coefficients $B_n^{SS}$ for constructing the steady state solution of the current step input for some selected normalized Faradaic conductance. Full data for $n \leq 1024$ and additional values of $G$ available online.

| | $G = 0$ | $G = 0.01$ | $G = 0.1$ | $G = 1$ | $G = 10$ | $G = 100$ |
|---|---|---|---|---|---|---|
| $B_1^{SS}$ | 1 | 1 | 1 | 1 | 1 | 1 |
| $B_2^{SS}$ | 0.31250 | 0.31183 | 0.30592 | 0.25912 | 0.11754 | 0.02564 |
| $B_3^{SS}$ | −0.05273 | −0.05296 | −0.05491 | −0.06795 | −0.07024 | −0.02418 |
| $B_4^{SS}$ | 0.01984 | 0.01991 | 0.02055 | 0.02610 | 0.04093 | 0.02142 |
| $B_5^{SS}$ | −0.00993 | −0.00996 | −0.01026 | −0.01291 | −0.02454 | −0.01849 |
| $B_6^{SS}$ | 0.00580 | 0.00581 | 0.00598 | 0.00746 | 0.01542 | 0.01573 |
| $B_7^{SS}$ | −0.00373 | −0.00374 | −0.00384 | −0.00476 | −0.01018 | −0.01328 |
| $B_8^{SS}$ | 0.00256 | 0.00257 | 0.00264 | 0.00325 | 0.00704 | 0.01118 |
| $B_9^{SS}$ | −0.00185 | −0.00186 | −0.00190 | −0.00234 | −0.00507 | −0.00939 |
| $B_{10}^{SS}$ | 0.00139 | 0.00139 | 0.00143 | 0.00175 | 0.00378 | 0.00790 |
| $B_{11}^{SS}$ | −0.00107 | −0.00108 | −0.00110 | −0.00135 | −0.00290 | −0.00666 |
| $B_{12}^{SS}$ | 0.00085 | 0.00085 | 0.00087 | 0.00107 | 0.00228 | 0.00563 |
| $B_{13}^{SS}$ | −0.00069 | −0.00069 | −0.00071 | −0.00086 | −0.00183 | −0.00478 |
| $B_{14}^{SS}$ | 0.00056 | 0.00057 | 0.00058 | 0.00071 | 0.00149 | 0.00408 |
| $B_{15}^{SS}$ | −0.00047 | −0.00047 | −0.00048 | −0.00059 | −0.00124 | −0.00350 |
| $B_{16}^{SS}$ | 0.00040 | 0.00040 | 0.00041 | 0.00050 | 0.00104 | 0.00302 |
| ⋮ | ⋮ | ⋮ | ⋮ | ⋮ | ⋮ | ⋮ |
| $V^{SS}/V_0$ | $+\infty$ | 101.08060 | 11.07922 | 2.06818 | 1.13327 | 1.01807 |

Table B5: Eigenvalues $\Lambda^{(i)}$ and coefficients $B_n^{(i)}$ for constructing the eigenfunctions of the transient response of current step input from basis functions. The trivial solution of $\Lambda^{(0)} = 0$ and $B_n^{(0)} = 0$, is excluded. Full data for $i, n \leq 1024$ available online.

| $i$ | 1 | 2 | 3 | 4 | 5 | 6 | 7 | 8 | 9 | 10 | ... |
|---|---|---|---|---|---|---|---|---|---|---|---|
| $\Lambda^{(i)}$ | 3.23686 | 5.76645 | 8.26009 | 10.74212 | 13.21888 | 15.69280 | 18.16501 | 20.63610 | 23.10643 | 25.57620 | ... |
| $B_1^{(i)}$ | 4.56973 | 3.77405 | 3.44403 | 3.25860 | 3.13835 | 3.05343 | 2.98996 | 2.94056 | 2.90091 | 2.86832 | ... |
| $B_2^{(i)}$ | 3.58511 | −3.70788 | −4.65165 | −4.79056 | −4.75592 | −4.67961 | −4.59717 | −4.51906 | −4.44807 | −4.38446 | ... |
| $B_3^{(i)}$ | 0.51738 | −7.51661 | −0.26793 | 2.76529 | 4.12700 | 4.79007 | 5.12852 | 5.30285 | 5.38887 | 5.42520 | ... |
| $B_4^{(i)}$ | 0.10883 | −2.89555 | 9.61985 | 5.38646 | 1.50528 | −1.06637 | −2.72102 | −3.80111 | −4.52140 | −5.01134 | ... |
| $B_5^{(i)}$ | −0.03142 | −0.67827 | 6.80910 | −8.19370 | −8.96686 | −6.36996 | −3.57332 | −1.25152 | 0.55192 | 1.92822 | ... |
| $B_6^{(i)}$ | 0.02274 | −0.02899 | 2.44679 | −10.7731 | 3.13093 | 8.87179 | 9.40950 | 7.93245 | 5.90871 | 3.91150 | ... |
| $B_7^{(i)}$ | −0.01587 | −0.03991 | 0.44950 | −5.76314 | 12.71019 | 3.84651 | −4.60764 | −8.76264 | −9.88246 | −9.36918 | ... |
| $B_8^{(i)}$ | 0.01161 | 0.02427 | 0.11317 | −1.72889 | 10.11847 | −10.9723 | −9.83824 | −2.30309 | 4.13705 | 8.05674 | ... |
| $B_9^{(i)}$ | −0.00879 | −0.01882 | −0.02225 | −0.42934 | 4.33792 | −14.0743 | 5.29853 | 12.04577 | 8.89869 | 2.86824 | ... |
| $B_{10}^{(i)}$ | 0.00684 | 0.01470 | 0.02444 | −0.02855 | 1.34838 | −8.32077 | 15.72582 | 2.78725 | −9.11201 | −12.08159 | ... |
| $B_{11}^{(i)}$ | −0.00544 | −0.01174 | −0.01900 | −0.03497 | 0.25080 | −3.31803 | 12.97479 | −13.5525 | −10.4004 | 1.88896 | ... |
| $B_{12}^{(i)}$ | 0.00441 | 0.00955 | 0.01553 | 0.02140 | 0.07855 | −0.89858 | 6.63451 | −16.8432 | 7.28305 | 14.42925 | ... |
| $B_{13}^{(i)}$ | −0.00364 | −0.00789 | −0.01285 | −0.01847 | −0.01763 | −0.23866 | 2.34443 | −11.13101 | 18.13177 | 1.66850 | ... |
| $B_{14}^{(i)}$ | 0.00304 | 0.00660 | 0.01078 | 0.01545 | 0.02133 | −0.01123 | 0.69463 | −4.95840 | 15.96042 | −15.44917 | ... |
| $B_{15}^{(i)}$ | −0.00257 | −0.00559 | −0.00914 | −0.01313 | −0.01740 | −0.02767 | 0.12123 | −1.73843 | 8.88252 | −19.64680 | ... |
| ⋮ | ⋮ | ⋮ | ⋮ | ⋮ | ⋮ | ⋮ | ⋮ | ⋮ | ⋮ | ⋮ | ⋱ |

Table B6: Coefficients $C^{(i)}$ of the eigenfunctions of the transient response of current step input for some selected normalized Faradaic conductance. Full data for $i \leq 128$ and additional values of $G$ available online.

| $n$ | $G = 0$ | $G = 0.01$ | $G = 0.1$ | $G = 1$ | $G = 10$ | $G = 100$ |
|---|---|---|---|---|---|---|
| $C^{(0)}$ | – | 100 | 10 | 1 | 0.1 | 0.01 |
| $C^{(1)}$ | 0.03692 | 0.03680 | 0.03582 | 0.02821 | 0.00903 | 0.00116 |
| $C^{(2)}$ | 0.01356 | 0.01354 | 0.01333 | 0.01156 | 0.00496 | 0.00074 |
| $C^{(3)}$ | 0.00710 | 0.00709 | 0.00702 | 0.00634 | 0.00321 | 0.00054 |
| $C^{(4)}$ | 0.00438 | 0.00438 | 0.00434 | 0.00401 | 0.00227 | 0.00042 |
| $C^{(5)}$ | 0.00298 | 0.00298 | 0.00296 | 0.00277 | 0.00170 | 0.00035 |
| $C^{(6)}$ | 0.00216 | 0.00216 | 0.00215 | 0.00203 | 0.00132 | 0.00029 |
| $C^{(7)}$ | 0.00164 | 0.00164 | 0.00163 | 0.00155 | 0.00106 | 0.00025 |
| $C^{(8)}$ | 0.00129 | 0.00128 | 0.00128 | 0.00123 | 0.00087 | 0.00022 |
| $C^{(9)}$ | 0.00104 | 0.00104 | 0.00103 | 0.00099 | 0.00072 | 0.00019 |
| $C^{(10)}$ | 0.00085 | 0.00085 | 0.00085 | 0.00082 | 0.00061 | 0.00017 |
| ⋮ | ⋮ | ⋮ | ⋮ | ⋮ | ⋮ | ⋮ |

**Voltage Step Response**

Table B7: Eigenvalues $\Lambda^{(i)}$ and coefficients $B_n^{(i)}$ for constructing the eigenfunctions of the transient response of voltage step input. Full data for $i, n \leq 1024$ available online.

| $i$ | 0 | 1 | 2 | 3 | 4 | 5 | 6 | 7 | 8 | 9 | ... |
|---|---|---|---|---|---|---|---|---|---|---|---|
| $\Lambda^{(i)}$ | 0.90931 | 3.39041 | 5.85921 | 8.32702 | 10.79460 | 13.26209 | 15.72954 | 18.19698 | 20.66440 | 23.13182 | ... |
| $B_0^{(i)}$ | 1 | 1 | 1 | 1 | 1 | 1 | 1 | 1 | 1 | 1 | ... |
| $B_1^{(i)}$ | 0.39451 | −3.30704 | −3.20144 | −3.08673 | −3.00260 | −2.94030 | −2.89258 | −2.85489 | −2.82433 | −2.79903 | ... |
| $B_2^{(i)}$ | −0.01974 | −3.09446 | 2.69232 | 3.87544 | 4.20749 | 4.29990 | 4.30959 | 4.28807 | 4.25489 | 4.21818 | ... |
| $B_3^{(i)}$ | 0.01259 | −0.52802 | 6.45944 | 0.65745 | −2.15584 | −3.53763 | −4.26428 | −4.66685 | −4.89691 | −5.02954 | ... |
| $B_4^{(i)}$ | −0.00657 | −0.10223 | 2.64610 | −8.32546 | −5.09803 | −1.69133 | 0.70001 | 2.29885 | 3.37463 | 4.11142 | ... |
| $B_5^{(i)}$ | 0.00393 | 0.02410 | 0.63787 | −6.16120 | 7.06425 | 8.21140 | 6.07128 | 3.56954 | 1.41585 | −0.29728 | ... |
| $B_6^{(i)}$ | −0.00256 | −0.01843 | 0.03554 | −2.27050 | 9.75696 | −2.49614 | −8.00976 | −8.76740 | −7.55547 | −5.74426 | ... |
| $B_7^{(i)}$ | 0.00178 | 0.01289 | 0.03502 | −0.43176 | 5.33232 | −11.52214 | −3.84585 | 3.99848 | 8.05051 | 9.27630 | ... |
| $B_8^{(i)}$ | −0.00129 | −0.00946 | −0.02056 | −0.10618 | 1.62963 | −9.36729 | 9.90614 | 9.29270 | 2.43707 | −3.64253 | ... |
| $B_9^{(i)}$ | 0.00097 | 0.00718 | 0.01605 | 0.01863 | 0.40730 | −4.07287 | 13.04355 | −4.63473 | −11.23179 | −8.54562 | ... |
| $B_{10}^{(i)}$ | −0.00075 | −0.00559 | −0.01255 | −0.02158 | 0.02998 | −1.27762 | 7.80961 | −14.57467 | −2.89008 | 8.37870 | ... |
| $B_{11}^{(i)}$ | 0.00060 | 0.00446 | 0.01003 | 0.01672 | 0.03202 | −0.24198 | 3.14134 | −12.18478 | 12.51967 | 9.96900 | ... |
| $B_{12}^{(i)}$ | −0.00048 | −0.00362 | −0.00817 | −0.01368 | −0.01915 | −0.07429 | 0.85897 | −6.28195 | 15.82526 | −6.60770 | ... |
| $B_{13}^{(i)}$ | 0.00040 | 0.00298 | 0.00675 | 0.01133 | 0.01661 | 0.01564 | 0.22817 | −2.23599 | 10.54522 | −17.02880 | ... |
| $B_{14}^{(i)}$ | −0.00033 | −0.00250 | −0.00565 | −0.00951 | −0.01390 | −0.01951 | 0.01206 | −0.66515 | 4.72715 | −15.12894 | ... |
| $B_{15}^{(i)}$ | 0.00028 | 0.00211 | 0.00479 | 0.00807 | 0.01181 | 0.01587 | 0.02580 | −0.11780 | 1.66486 | −8.47037 | ... |
| ⋮ | ⋮ | ⋮ | ⋮ | ⋮ | ⋮ | ⋮ | ⋮ | ⋮ | ⋮ | ⋮ | ⋱ |

Table B8: Coefficients $C^{(i)}$ of the eigenfunctions of the transient response of voltage step input for some selected normalized Faradaic conductance. Full data for $i \leq 128$ and additional values of $G$ available online.

| $n$ | $G = 0$ | $G = 0.01$ | $G = 0.1$ | $G = 1$ | $G = 10$ | $G = 100$ |
|---|---|---|---|---|---|---|
| $C^{(0)}$ | −0.88889 | −0.87922 | −0.80082 | −0.42334 | −0.07409 | −0.87922 |
| $C^{(1)}$ | −0.05700 | −0.05683 | −0.05537 | −0.04402 | −0.01443 | −0.05683 |
| $C^{(2)}$ | −0.01865 | −0.01862 | −0.01834 | −0.01593 | −0.00689 | −0.01862 |
| $C^{(3)}$ | −0.00914 | −0.00913 | −0.00903 | −0.00816 | −0.00415 | −0.00913 |
| $C^{(4)}$ | −0.00541 | −0.00540 | −0.00536 | −0.00495 | −0.00281 | −0.00540 |
| $C^{(5)}$ | −0.00357 | −0.00357 | −0.00354 | −0.00332 | −0.00203 | −0.00357 |
| $C^{(6)}$ | −0.00253 | −0.00253 | −0.00251 | −0.00238 | −0.00155 | −0.00253 |
| $C^{(7)}$ | −0.00189 | −0.00189 | −0.00188 | −0.00179 | −0.00122 | −0.00189 |
| $C^{(8)}$ | −0.00146 | −0.00146 | −0.00145 | −0.00139 | −0.00098 | −0.00146 |
| $C^{(9)}$ | −0.00116 | −0.00116 | −0.00116 | −0.00112 | −0.00081 | −0.00116 |
| $C^{(10)}$ | −0.00095 | −0.00095 | −0.00095 | −0.00091 | −0.00068 | −0.00095 |
| ⋮ | ⋮ | ⋮ | ⋮ | ⋮ | ⋮ | ⋮ |

# Bibliography

Newman, J. Resistance for Flow of Current to a Disk. Journal of the Electrochemical Society, 113(5), pp. 501–502, 1966a.

Newman, J. Current Distribution on a Rotating Disk below the Limiting Current. Ibid., 113(12), pp. 1235–1241, 1966b.

Newman, J. The Diffusion Layer on a Rotating Disk Electrode. Ibid., 114(3), p. 239, 1967.

Marathe, V. and Newman, J. Current Distribution on a Rotating Disk Electrode. Journal of the Electrochemical Society, 116(12), pp. 1704–1707, 1969.

Newman, J. Ohmic Potential Measured by Interrupter Techniques. Ibid., 117(4), pp. 507–508, 1970a.

Newman, J. Frequency Dispersion in Capacity Measurements at a Disk Electrode. Ibid., 117(2), pp. 198–203, 1970b.

Smyrl, W. H. and Newman, J. Limiting Current on a Rotating Disk with Radial Diffusion. Ibid., 118(7), pp. 1079–1081, 1971.

Smyrl, W. H. and Newman, J. Detection of Nonuniform Current Distribution on a Disk Electrode. Ibid., 119(2), pp. 208–212, 1972.

Tiedemann, W. H., Newman, J., and Bennion, D. N. The Error in Measurements of Electrode Kinetics Caused by Nonuniform Ohmic-Potential Drop to a Disk Electrode. Ibid., 120(2), pp. 256–258, 1973.

Nisancioğlu, K. and Newman, J. The Transient Response of a Disk Electrode. Ibid., 120(10), pp. 1339–1346, 1973a.

Nisancioğlu, K. and Newman, J. The Transient Response of a Disk Electrode with Controlled Potential. Ibid., 120(10), pp. 1356–1358, 1973b.

Vahdat, N. and Newman, J. *Corrosion of an Iron Rotating Disk.* Ibid., 120(12), pp. 1682–1686, 1973.

Homsy, R. V. and Newman, J. An Asymptotic Solution for the Warburg Impedance of a Rotating Disk Electrode. Ibid., 121(4), pp. 521–523, 1974a.

Nisancioğlu, K. and Newman, J. The Short-Time Response of a Disk Electrode. Ibid., 121(4), pp. 523–527, 1974.

Homsy, R. V. and Newman, J. Current Distribution on a Plane below a Rotating Disk. Ibid., 121(11), pp. 1448–1451, 1974b.

Pierini, P., Appel, P., and Newman, J. Current Distribution on a Disk Electrode for Redox Reactions. Ibid., 123(3), pp. 366–369, 1976.

Pierini, P. and Newman, J. Potential Distribution for Disk Electrodes in Axisymmetric Cylindrical Cells. Ibid., 126(8), pp. 1348–1352, 1979.

Tribollet, B. and Newman, J. Analytic Expression of the Warburg Impedance for a Rotating Disk Electrode. Ibid., 130(4), pp. 822–824, 1983.

Law, C. G. and Newman, J. Corrosion of a Rotating Iron Disk in Laminar, Transition, and Fully Developed Turbulent Flow. Ibid., 133(1), pp. 37–42, 1986.

Jakšić, M. M. and Newman, J. The Kramers-Kronig Relations and Evaluation of Impedance for a Disk Electrode. Ibid., 133(6), pp. 1097–1101, 1986.

Smyrl, W. H. and Newman, J. Current Distribution at Electrode Edges at High Current Densities. Ibid., 136(1), pp. 132–139, 1989.

West, A. C. and Newman, J. Corrections to Kinetic Measurements Taken on a Disk Electrode. Ibid., 136(1), pp. 139–143, 1989a.

West, A. C. and Newman, J. Current Distribution near an Electrode Edge as a Primary Distribution Is Approached. Ibid., 136(10), pp. 2935–2939, 1989b.

Li, S. X.-Z. and Newman, J. Cathodic Protection for Disks of Various Diameters. Ibid., 148(4), pp. B157–B162, 2001.

Wiley, J. D. and Webster, J. G. Analysis and Control of the Current Distribution under Circular Dispersive Electrodes. IEEE Transactions on Biomedical Engineering, BME-29(5), pp. 381–385, 1982.

Boyd, J. P. Chebyshev and Fourier spectral methods, Second Edition (Revised). Dover Publications, Inc., Mineola, NY, USA, 2001.

A. Richardot and E. T. McAdams. Harmonic analysis of low-frequency bioelectrode behavior. IEEE Transactions on Medical Imaging, 21(6), pp. 604–612, 2002.

Oldham, K. B. The RC time "constant" at a disk electrode. Electrochemical Communication, 6(2), pp. 210–214, 2004.

Newman, J. Electrochemical systems. 3rd ed., Englewood Cliffs, NJ: Prentice Hall, 2004

Myland, J. C. and Oldham, K. B. How does the double layer at a disk electrode charge? Journal of Electroanalytical Chemistry, 575(1), pp. 81–93, 2005.

Antohi, P. and Scherson, D. A, Current Distribution at a Disk Electrode during a Current Pulse, Journal of The Electrochemical Society 153 (2), pp. E17–E24, 2006.

Huang, V. M.-W., Vivier, V., Orazem, M. E., Pébère, N., and Tribollet, B. The Apparent Constant-Phase-Element Behavior of an Ideally Polarized Blocking Electrode A Global and Local Impedance Analysis. Journal of The Electrochemical Society, 154(2), pp. C81–C88, 2007a.

Huang, V. M.-W., Vivier, V., Orazem, M. E., Pébère, N., and Tribollet, B. The Apparent Constant-Phase-Element Behavior of a Disk Electrode with Faradaic Reactions A Global and Local Impedance Analysis. Journal of The Electrochemical Society, 154(2), pp. C99–C107, 2007b.

Orazem, M. E. and Tribollet B., Electrochemical Impedance Spectroscopy. John Wiley & Sons: Hoboken, NJ, USA, 2008.

Orazem, M. E. and Tribollet, B. Perspectives on Newman's Work on Resistance for Flow of Current to a Disk. The Electrochemical Society Interface, pp. 56–58, Spring 2009.

Wang, B., Petrossians, A., and Weiland, J. D. Reduction of Edge Effect on Disk Electrodes by Optimized Current Waveform. IEEE Transactions on Biomedical Engineering, 61(8), pp. 2254–2263, 2014.

Wang, B. Investigation of the Electrode-Tissue Interface of Retinal Prostheses. Department of Biomedical Engineering, University of Southern California, Los Angeles, CA, USA, May 2016. ProQuest Dissertation No. 10124439.

Newman, J. and Battaglia, V. The Newman Lectures on Mathematics. Jenny Stanford Publishing, New York, NY, USA, 2018.

Chen, Z., Ryzhik, L., and Palanker, D. Current Distribution on Capacitive Electrode-Electrolyte Interfaces. Physical Review Applied, 13(1), p. 014004, 2020.